\documentclass[pre,showpacs,preprintnumbers,twocolumn,amsmath,amssymb,superscriptaddress]{revtex4-1}
\usepackage{bm}
\usepackage{graphicx,color}
\usepackage{graphicx}
\usepackage{dcolumn}
\usepackage{times}
\UseRawInputEncoding

\usepackage{amsmath,amssymb}

\begin{document}

\title{Power laws of natural swarms are fingerprints of an extended critical region}
\author{R. Gonz\'alez-Albaladejo}
\affiliation{Departamento de Matem\'atica Aplicada, Universidad Complutense de Madrid, 28040 Madrid, Spain}
\affiliation{Gregorio Mill\'an Institute for Fluid Dynamics, Nanoscience and Industrial Mathematics, Universidad Carlos III de Madrid, 28911 Legan\'{e}s, Spain}
\author{L. L. Bonilla$^*$}
\affiliation{Gregorio Mill\'an Institute for Fluid Dynamics, Nanoscience and Industrial Mathematics, Universidad Carlos III de Madrid, 28911 Legan\'{e}s, Spain}
\affiliation{Department of Mathematics, Universidad Carlos III de Madrid, 28911 Legan\'{e}s, Spain. 
$^*$Corresponding author. E-mail: bonilla@ing.uc3m.es}
\date{\today}
\begin{abstract}
Collective biological systems display power laws for macroscopic quantities and are fertile probing grounds for statistical physics. Besides power laws, natural insect swarms present strong scale-free correlations, suggesting closeness to phase transitions. Swarms exhibit {\em  imperfect} dynamic scaling: their dynamical correlation functions collapse into single curves when written as functions of the scaled time $t\xi^{-z}$ ($\xi$: correlation length, $z$: dynamic exponent), but only for short times. Triggered by markers, natural swarms are not invariant under space translations. Measured static and dynamic critical exponents differ from those of equilibrium and many nonequilibrium phase transitions. Here, we show that: (i) the recently discovered scale-free-chaos phase transition of the harmonically confined Vicsek model has a novel extended critical region for $N$ (finite) insects that contains several critical lines. (ii) As alignment noise vanishes, there are power laws connecting critical confinement and noise that allow calculating static critical exponents for fixed $N$. These power laws imply that the unmeasurable confinement strength is proportional to the perception range measured in natural swarms. (iii) Observations of natural swarms occur at different times and under different atmospheric conditions, which we mimic by considering mixtures of data on different critical lines and $N$. Unlike results of other theoretical approaches, our numerical simulations reproduce the previously described features of natural swarms and yield static and dynamic critical exponents that agree with observations.
\end{abstract}

\maketitle

\section{Introduction}\label{sec:1}
The formation of animal flocks presents common features irrespective of biological details \cite{oku86,oku01,par99,yat09,sum10,vic12,str13,gin15,oue22} and it is a precursor of the major transitions in the evolution of complexity \cite{hux12,smi95} (e.g.,  changes from single cell protists to multicellular organisms, changes from individual ants, bees and other insects to their society \cite{hux12,smi95}). In particular, many macroscopic observables of biological systems obey power laws and their critical exponents have been measured \cite{mor11,bia12,ple14,tan17,zam22,sum10,aza18,cav18}. Mitochondrial networks \cite{zam22}, bacterial colonies \cite{zha10}, bird flocks \cite{bal08,cav10,bia12pnas} and insect swarms \cite{att14,cav17} provide examples of scale free behavior as their correlation length increases with the size of the flock, thereby rendering irrelevant intrinsic length scales associated to individuals \cite{cav18}. Since the scale free property accompanies phase transitions, there have been many theoretical studies on the possible phase transitions responsible for flocking and other collective behavior in dry active matter \cite{cha20}, starting with the works by Vicsek {\em et al} \cite{vic95} and Toner and Tu \cite{ton95}. 

The interaction between swarming midges is acoustic and insects interact when their distances are sufficiently small \cite{att14}. The distribution of midge speeds in a swarm is peaked about some value with heavy tails for large swarms (perhaps due to the formation of clusters) \cite{kel13}. The statistics of accelerations of individual midges in a swarm is consistent with postulating a linear spring force (therefore a harmonic potential) that binds insects together \cite{kel13}. Swarm of midges in the wild exhibit long range correlations \cite{att14plos,att14,cav17,cav23}, which are absent in laboratory conditions without background noise and atmospheric variability \cite{ni15epj}. Here we are interested in power laws for correlation length, time and susceptibility of natural swarms, which are associated with strong correlations. Thus, we adopt the Vicsek model (VM) metric alignment of an insect with neighbors within a sphere of influence as a reasonable choice, and ignore variations in the individual speed. We also include a linear spring force to confine the swarm \cite{oku86,kel13,gor16}. The effects of a fluctuating speed could be the subject of future studies. 

For starling flocks \cite{bal08}, neighbors are topologically defined, metric-free models may incorporate a distributed motional bias \cite{lew17}, bird rotations propagate swiftly as linear waves \cite{cav18} and a reasonable extension of the continuous-time VM is the inertial spin model \cite{cav15}. Furthermore, visual and auditory sensing are compared in \cite{roy19}, the influence of time delay is studied in \cite{gei22}, the influence of metric and topological interactions on flocking is studied in \cite{kum21} and \cite{rey17} considers a swarming model based on effective velocity-dependent gravity. VM based in social interactions do not account for features of bird flocks and fish schools based on vortices shed by flapping. Hydrodynamic interactions theoretically studied in \cite{oza19} are important for observed ordered schools and bird flocks. Noise may induce schooling in finitely many fish experiencing binary interactions \cite{jha20}. A modified VM with varying speeds and infinite circle of influence exhibits transitions between migrating and rotating states of the fish school \cite{bir07}.  

An attractive feature of the phase transition analogy is the notion of {\em universality}: different models belonging to the same universality class have the same scale-free limit and critical exponents determined by renormalization group (RG) flow \cite{wil74}. Dynamics complicates this picture: different dynamic laws may produce the same static critical exponents but different dynamic critical exponents about an equilibrium phase transition \cite{hoh77}. Assuming universality, calculations on simple models can be compared to measurements of critical exponents of biological systems. This is Cavagna {\em et al}'s point of view in their study of midge swarms \cite{cav23}: {\em Assuming that swarms are close to an ordering transition between homogeneous phases}, the RG flow for sufficiently rich dynamics produces a dynamical critical exponent close to the observed one. They explore the RG flow of an active model of types E/F and G in Ref.~\cite{hoh77} and simulate numerically the inertial spin model with periodic boundary conditions \cite{cav15}.

Nonetheless, the origin of power laws in observations of insect swarms remains puzzling. As mentioned before, the long range correlations observed in natural swarms \cite{att14plos,att14,cav17,cav23} are absent in laboratory conditions without background noise and atmospheric variability \cite{ni15epj}. Measurements of the critical dynamical exponent $z$ between the correlation time $\tau$ and the correlation length $\xi$ produce values in a range between $z=1.16$ and 1.37 depending on sampling and fitting procedures \cite{cav17,cav23}. When written in terms of the scaled time $t/\xi^z$ and measured on natural swarms, the normalized dynamic connected correlation function (NDCCF) collapses into a single curve only on a finite interval (approximately $0<t/\xi^z<4$) \cite{cav17}. However, the same function collapses for all scaled times according to theories based on standard RG ideas \cite{ami05,hoh77} for the ordering phase transition of a complex system of stochastic partial differential equations (PDEs) \cite{cav23}. Measured and predicted $z$ values are very close but static critical exponents are not \cite{cav23}. Homogeneous phases in the ordering transition are invariant under space translations. Natural swarms are not, because they form over specific darker spots on the ground (wet areas, cow dung, man-made objects, etc) called markers \cite{dow55}.

What is going on? Simply put, natural swarms are not close to an {\em ordering phase transition} and their `temperature' (whatever acts as control parameter) is different during different observations. Here we propose an alternative theory based on a number of technical discoveries for the harmonically confined VM (HCVM). The HCVM involves a number of simplifications (equal speed and isotropy of insect velocities), while real swarms have speeds distributed about a maximum value and vertical velocities are smaller than horizontal ones \cite{kel13}. More refined models may be explored on the basis of the present study. In a nutshell, the HCVM has a scale-free-chaos phase transition with an extended criticality region on parameter space whose critical lines collapse at the same rate as the number of insects $N$ goes to infinity. Observations of natural swarms sample the extended criticality region for different $N$ and values of the control parameter. Using the same methodology as in observations, we obtain dynamic and static critical exponents close to those measured. Moreover, the NDCCF collapses into a single curve for $0<t/\xi^z<4$. 

{\em Harmonically confined Vicsek model.} For finite $N$, the three dimensional HCVM on the plane $(\eta,\beta)$ is:
\begin{eqnarray}
&&\mathbf{x}_i(t+1)=\mathbf{x}_i(t)+ \mathbf{v}_i(t+1),\quad i=1,\ldots,N,\nonumber\\
&& \mathbf{v}_i(t+1)=v_0  \mathcal{R}_\eta\!\left[\Theta\!\left(\sum_{|\mathbf{x}_j-\mathbf{x}_i|<R_0}\mathbf{v}_j(t)-\beta\mathbf{x}_i(t)\right)\!\right]\!. \label{eq1}
\end{eqnarray}
Here $\Theta(\mathbf{x})=\mathbf{x}/|\mathbf{x}|$, $R_0$ is the radius of the sphere of influence about particles, $v_0$ is the constant particle speed, $\beta$ is the confining spring constant, and $\mathcal{R}_\eta(\mathbf{w})$ performs a random rotation uniformly distributed on a spherical sector around $\mathbf{w}$ with maximum opening $\eta$ \cite{gon23}. Particles align their velocities with the mean of their neighbors within a sphere of influence except for an alignment noise of strength $\eta$.

Technical discoveries are as follows. Firstly, the HCVM exhibits a phase transition characterized by scale-free chaos and an extended criticality region \cite{gon23,gon23mf}. There are three critical lines on the noise-confinement phase plane $(\eta,\beta)$ having $\xi\sim N^\frac{1}{3}$ that collapse at the same rate to the $\beta=0$ axis as the insect number $N\to\infty$: the single-to-multicluster chaos line, $\beta_c(N,\eta)$, the line of maximal largest Lyapunov exponents (LLE), $\beta_i(N,\eta)$, and the onset of chaos line (zero LLE), $\beta_0(N,\eta)$ \cite{gon23,gon23mf}. For finite $N$, the region comprising these lines is an extended criticality region. Secondly, as the noise $\eta\to 0$, there are power laws connecting critical confinement to $N$ and $\eta$ that involve the critical static exponent $\nu$. These power laws can be used to estimate $\nu$ at fixed $N$, a valuable result because the insect number cannot be increased at will. Thirdly, the same power laws with noise imply that the measurable perception range (time averaged arithmetic mean of the minimal distance between each insect and its closest neighbor \cite{att14}) is proportional to confinement on the critical lines, a control parameter that cannot be directly measured. Lastly, the value of the dynamic critical exponent depends on how it is measured. When the HCVM is simulated for different $N$, $\eta$ and $\beta$ within the criticality region on lines $\beta_0$ and $\beta_c$ (mimicking experimental conditions), we obtain $z=1.15\pm 0.11$ using least squares (LS) fitting and $z=1.33\pm 0.10$ by reduced major axis (RMA) regression \cite{cav23}. However, both methods produce the same values when calculated on the same critical line at fixed $\eta$ and variable $N$, e.g., $z=1.01\pm0.01$ on $\beta_c$ \cite{gon23}.

The rest of the paper is as follows. The phase diagram of different phases on regions of the plane $(\eta,\beta)$ is studied in Section \ref{sec:2}. There are three critical lines that tend to zero confinement as $N\to\infty$ at the same rate. These lines define an extended criticality region of the phase diagram on which scale-free behavior is expected. These lines issue forth from the origin $\eta=\beta=0$ at finite $N$ as power laws of $\beta$ in terms of the noise. These power laws can be used to deduce the static critical exponent $\nu$ using data at a single $N$. Section \ref{sec:3}  recalls the definition of static and dynamic connected correlation functions, correlation length and correlation times, as well as the dynamic scaling hypothesis and the definitions of the critical exponents, which are calculated on the different critical lines. We also show that the confinement control parameter is related to the perception range, which can be measured in natural swarms. Then the power laws defining critical exponents can be expressed in terms of the perception range, as it was done in Ref.~\cite{att14} for the static critical exponents. Section \ref{sec:4} considers mixtures of data in the extended critical region as a reasonable model for the experimental data obtained from natural swarms. We show that the dynamical correlation function data collapse when time is scaled as $t\xi^{-z}$ on an interval of finite length that begins at $t=0$.  We also calculate the dynamic critical exponent by least square and by reduced major axis regressions and show that the obtained values are close to those measured in natural swarms. Section \ref{sec:5} discusses our results and contains our conclusions. Appendix \ref{ap:a} (adapted from \cite{gon23} with slight modifications) explains the algorithms used to calculate the largest Lyapunov exponent and to reconstruct attractors from time series.

\begin{widetext}
\begin{center}
\begin{figure}[ht]
\begin{center}
\includegraphics[clip,width=18cm]{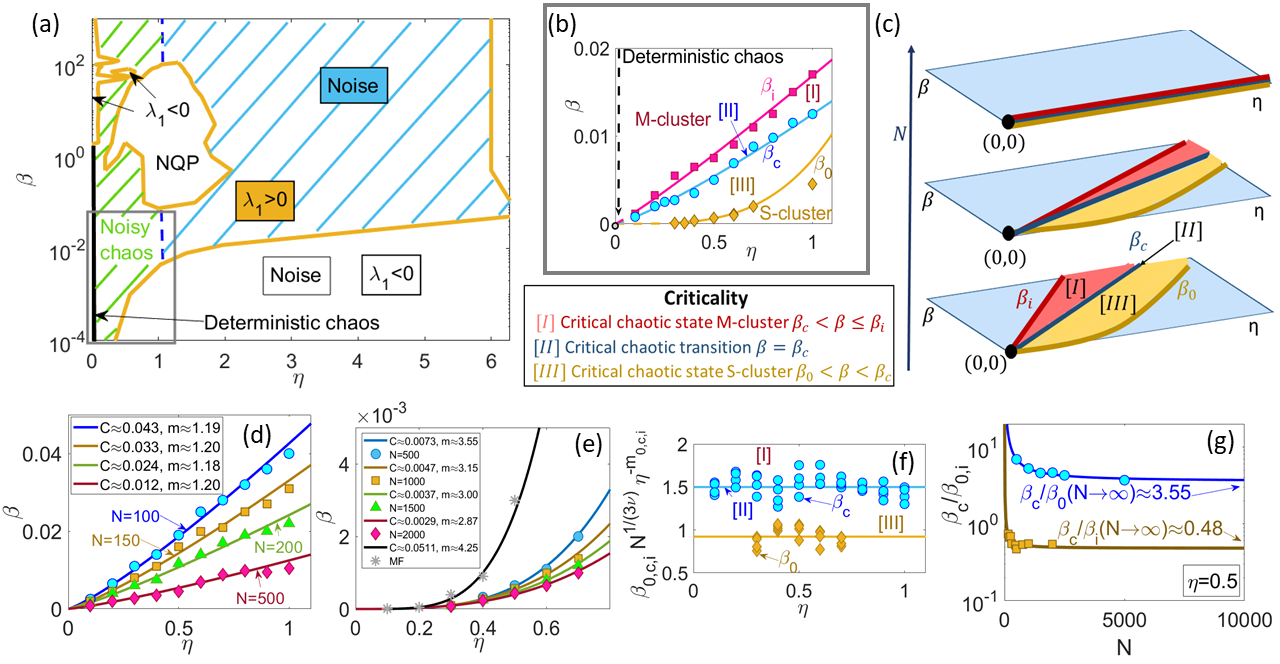}\\
\end{center}
\caption{{\bf (a)} Phase diagram on confinement vs noise plane for $N=500$, $v_0=R_0=1$, indicating regions of deterministic and noisy chaos, noisy period-$\sigma$ (NP$\sigma$) and noisy quasiperiodic (NPQ) attractors, and mostly noise. {\bf (b)} For $N=500$, regions [I]  $(\beta_c(\eta;N),\beta_i(\eta;N))$ (M-cluster or multicluster chaos), [III] $(\beta_0(\eta;N),\beta_c(\eta;N))$ (S-cluster or single-cluster chaos), and line [II] $\beta=\beta_c(\eta;N)$. {\bf (c)} Shrinking of the criticality region as $N$ increases. {\bf (d)} $\beta_c(\eta;N)\sim C_cN^{-1/(3\nu)}\eta^{m_c}$, $C_c=1.5\pm 0.2$,  $m_c= 1.20\pm 0.04$. {\bf (e)}  Curve $\beta_0(N;\eta)\sim C_0N^{-1/(3\nu)}\eta^{m_0}$, $C_0=0.92\pm 0.22$, $m_0\sim m_c+a_2N^{-n_2}$, $a_2=2.36\pm 0.07$, $n_2=0.24\pm0.01$, separating chaotic and non-chaotic regions. We have also indicated the mean-field  approximate curve \cite{gon23mf}. {\bf (f)} Collapse of curves of panels (d), (e): $\beta_{0,c}(\eta;N) N^{1/(3\nu)}\eta^{-m_{0,c}}$ vs $\eta$. {\bf (g)} Numerical illustration of the relations $\beta_c/\beta_0\approx 3.55$ , $\beta_c/\beta_i\approx 0.48$ as $N\to\infty$ for $\eta=0.5$.   \label{fig1}}
\end{figure}
\end{center}
\end{widetext}

\section{Phase diagram and critical lines} \label{sec:2}
For finite $N$, Figure \ref{fig1}(a) depicts the phase diagram of the three dimensional HCVM on the plane $(\eta,\beta)$. It resembles the phase diagram of the mean-field (MF) equation for the swarm center of mass, given by Eq.~\eqref{eq1} with $N=1$ \cite{gon23mf}. Numerical simulations of the HCVM show that there are regions in the parameter space where the largest Lyapunov exponent is positive, indicating the existence of chaotic attractors. There is a narrow region of deterministic chaos close to $\eta=0$, for larger $\eta$, another region corresponds to noisy chaos followed by a larger region where noise swamps chaos. See Appendix \ref{ap:a} and Ref.~\cite{gon23} for the technical definition of noisy chaos using scale-dependent Lyapunov exponents, calculations of the LLE and reconstruction of chaotic attractors from time series obtained from the numerical simulations of the HCVM. The region of positive LLE are bounded by different curves. For sufficiently large confinement, Fig.~\ref{fig1}(a) shows regions of noisy quasiperiodic attractors. For sufficiently small $\beta$, the LLE is non-positive and attractors are non-chaotic. On the curve $\beta_0(N;\eta)$ separating chaotic and nonchaotic attractors for fixed $N$, the LLE is zero. On this curve the correlation length defined below is proportional to the size of the swarm and all other length scales are irrelevant, which indicates scale-free behavior and characterizes the phase transition in the limit as $N\to\infty$.

\begin{center}
\begin{figure}[ht]
\begin{center}
\includegraphics[clip,width=4.2cm]{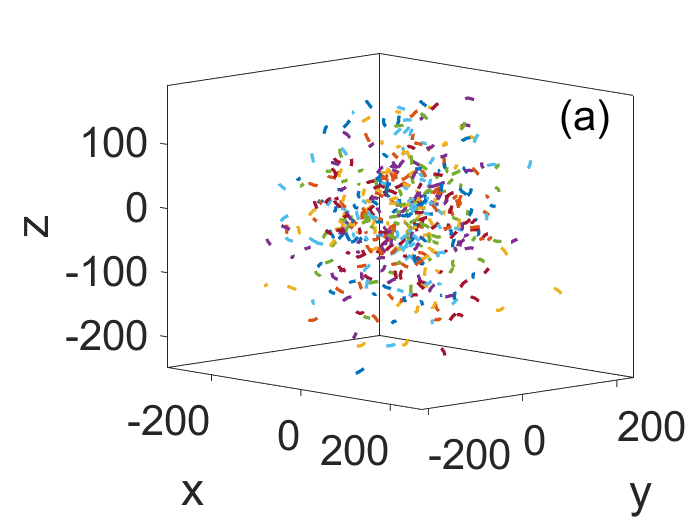}
\includegraphics[clip,width=4.2cm]{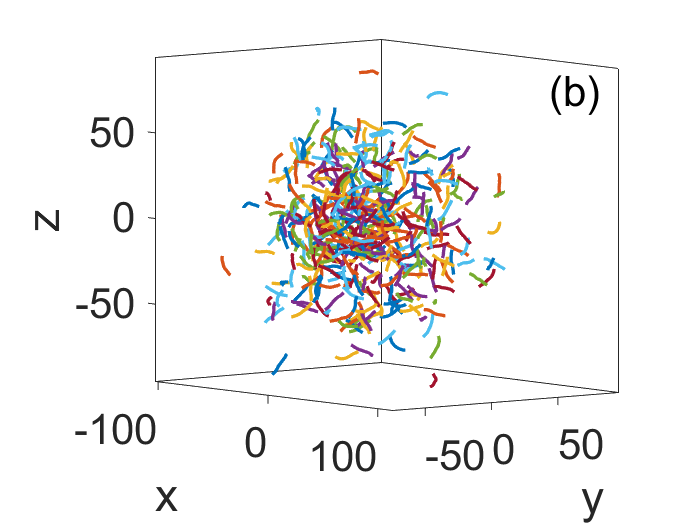}\\
\includegraphics[clip,width=4.2cm]{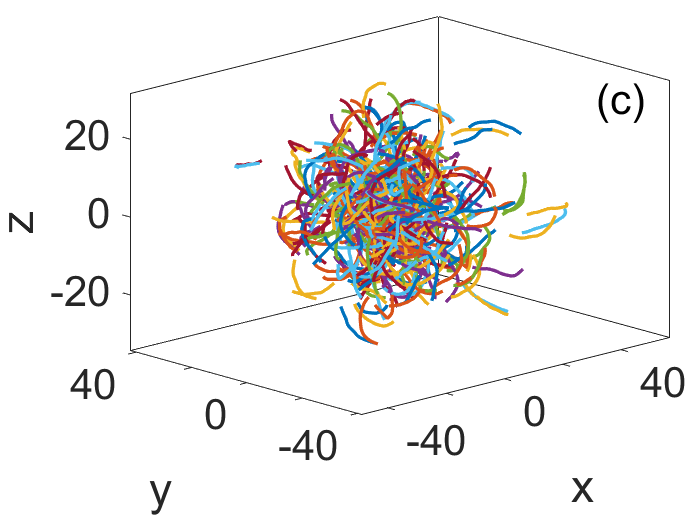}
\includegraphics[clip,width=4.2cm]{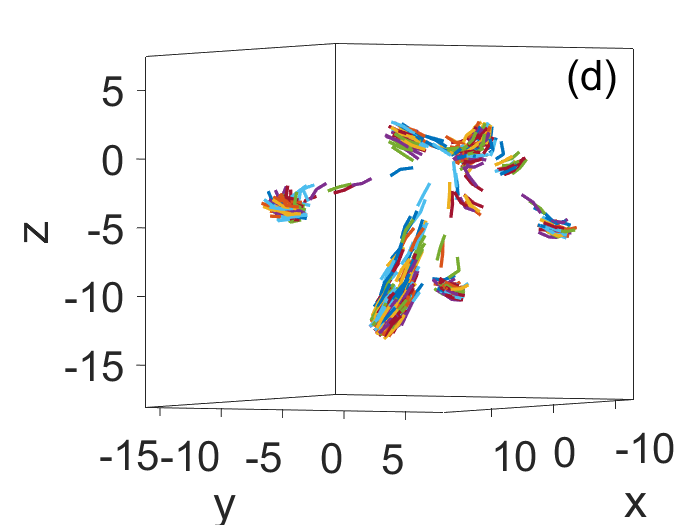}
\\
\end{center}
\caption{Swarm as depicted from short trajectories of 500 particles and $\eta=0.5$ for confinements {\bf (a)} $\beta=0.0001$, {\bf (b)} $\beta_0=0.00065$,  {\bf (c)} $\beta_c=0.005$, {\bf (d)} $\beta=1$. See associated videos in the Supplementary Material \cite{suppl}.}
 \label{fig2}
\end{figure}
\end{center}

Fig.~\ref{fig1}(b) shows three scale-free lines where swarm size and correlation length are proportional, $\beta_0<\beta_c<\beta_i$. $\beta_0(\eta;N)$ separates regions of chaotic attractors (LLE $\lambda_1>0$ in [III]) from nonchaotic regions (negative LLE) \cite{gon23mf}; $\beta_c(\eta;N)$ (line [II]) separates chaotic single from multicluster swarms in region [I], whereas the LLE are maximal on the line $\beta_i(\eta;N)$ \cite{gon23}. Fig.~\ref{fig2} shows the shape of the swarm for increasing values of $\beta$ as deduced from depicting short trajectories of its particles. The scale-free lines $\beta_c$ and $\beta_i$ were found in Ref.~\cite{gon23}, whereas the scale-free line $\beta_0$ was studied for the MF HCVM in Ref.~\cite{gon23mf}. 

Fig.~\ref{fig1}(c) shows that the extended criticality region between $\beta_0(\eta;N)$ and $\beta_i(\eta;N)$ shrinks with increasing $N$. The three critical lines collapse into the noise axis at the same rate as $N\to\infty$; see Fig.~\ref{fig1}(g). Finite-size and dynamical scaling imply $\xi\sim\beta^{-\nu}$, $\chi\sim\beta^{-\gamma}$ (susceptibility), $\tau\sim\xi^z$ (correlation time) \cite{gon23}. Fig.~\ref{fig1}(d) and \ref{fig1}(e) illustrate how $\beta_c$ and $\beta_0$, respectively, depend on $\eta$ for different $N$. In rescaled coordinates, these curves collapse for fixed $\eta$ as shown in Fig.\! \ref{fig1}(f):  
\begin{eqnarray}
\beta_j(N;\eta)= C_j N^{-\frac{1}{3\nu}}\eta^{m _j}, \quad j=0,c, \label{eq2}
\end{eqnarray}
where $C_c=1.5\pm 0.2$, $m_c=1.2\pm 0.04$, $C_0=0.92\pm 0.22$, $m_0=m_c+a_2 N^{-n_2}$, $m_c= 1.20\pm 0.04$, $a_2=2.36 \pm0.07$, $n_2=0.24\pm0.01$. As the static critical exponents {\em are independent of} $\eta$ \cite{gon23}, the power laws \eqref{eq2} allow calculating $\nu=0.43\pm 0.03$  and $\gamma=0.92\pm 0.13$ using one or several values of $N$; see also Section \ref{sec:3}.  This is {\em a major result} because the critical exponents are found from power laws in $\eta$ without resorting to numerical simulations for ever increasing particle numbers. Whether this also occurs for other space dimensions or phase transitions is matter for future research.
\bigskip

\section{Critical curves and critical exponents for finite $N$ as $(\beta,\eta)\to (0,0)$ }\label{sec:3}
In MF theory, the zero noise and confinement limits correspond to the scale-free-chaos phase transition and the correlation length, time, order parameter and susceptibility have to be defined in terms of the swarm center-of-mass motion \cite{gon23mf}. Then $\xi=\langle R\rangle_t$ or $\xi=$ max$R$, with $R(t)=|\mathbf{X}(t)|$, $1/\tau=w=\Omega$ ($w$ and $\Omega$ are the winding number and the maximum frequency of the spectrum for the time series $X(t)+Y(t)+Z(t)$, respectively), $w=\Omega$ plays the role of order parameter, and the susceptibility is defined by linear response to an external force $\mathbf{H}$ added to Eq.~\eqref{eq1} \cite{gon23mf}.  We now recall the definitions of static and dynamic connected correlation functions (SCCF and DCCF, respectively), correlation length and correlation time. Then we describe our results for the different critical lines.

\subsection{Correlation functions}
\begin{widetext}
The DCCF is  \cite{att14,cav18}
\begin{eqnarray}\
&&C(r,t)\!=\!\!\left\langle\! \frac{\sum_{i=1}^{N}\!\sum_{j=1}^{N}\delta\hat{\mathbf{v}}_i(t_0\!)\!\cdot\!\delta\hat{\mathbf{v}}_j(t_0\!+t)\delta[r\!-r_{ij}(t_0\!,t)]}{\sum_{i=1}^{N}\sum_{j=1}^{N}\delta[r-r_{ij}(t_0,t)]}\! \right\rangle_{t_0}\quad \label{eq3}\\
&&C(r)=C(r,0),\nonumber\\
&&\delta\hat{\mathbf{v}}_i\!=\frac{\delta\mathbf{v}_i}{\sqrt{\frac{1}{N}\sum_k \delta\mathbf{v}_k\cdot\delta\mathbf{v}_k}},\quad \delta\mathbf{v}_i=\mathbf{v}_i - \mathbf{V},\nonumber\\
&&r_{ij}(t_0,t)=|\mathbf{r}_i(t_0)-\mathbf{r}_j(t_0+t)|,\,  \mathbf{r}_i(t_0)=\mathbf{x}_i(t_0) - \frac{1}{N}\sum_{j=1}^N\mathbf{x}_j(t_0),   \nonumber\\
&&\langle f\rangle_{t_0}=\frac{1}{t_{max}-t}\sum_{t_0=1}^{t_{max}-t}f(t_0,t).\nonumber 
\end{eqnarray}
In these equations, $\delta(r-r_{ij})=1$ if $r<r_{ij}<r+dr$ and zero otherwise, and $dr$ is the space binning factor. The averages are over time and over five independent realizations corresponding to five different random initial conditions during 10000 iterations \cite{gon23}. The SCCF is the equal time connected correlation function $C(r)=C(r,0)$ given by Eq.~\eqref{eq3}. Note that $C(\infty)\propto |\sum_{i=1}^N\delta\mathbf{\hat{v}}_i|^2=0$. The correlation length $\xi$ can be defined as the first zero of $C(r)$, $r_0$, corresponding to the first maximum of the cumulative correlation function \cite{att14}:
\begin{eqnarray}
&&Q(r)=\!\left\langle \frac{1}{N}\sum_{i=1}^{N}\sum_{j=1}^{N} \delta\hat{\mathbf{v}}_i\!\cdot\!\delta\hat{\mathbf{v}}_j\theta(r-r_{ij}(t_0,0))\right\rangle_{t_0}\!, \quad\chi= Q(\xi),\label{eq4}\\ 
&&\xi=\mbox{argmax}Q(r),\, C(\xi)=0\,\mbox{ with }\, C(r)>0,\, r\in(0,\xi), \quad\nonumber
\end{eqnarray}
where $\theta(x)$ is the Heaviside unit step function. For $r$ larger than the swarm size, $Q(r)=\langle |\sum_{i=1}^N \delta\mathbf{\hat{v}}_i|^2 \rangle_{t_0}/N=0$. The susceptibility $\chi$ is the value of $Q(r)$ at its first maximum, as in Ref.~\onlinecite{att14}. Alternatively, we can use the Fourier transform of Eq.~\eqref{eq3},
\end{widetext}
\begin{eqnarray}
\hat{C}(k,t)\!=\!\left\langle\! \frac{1}{N}\!\sum_{i,j =1}^{N}\!\!\frac{\sin(kr_{ij}(t_0,t))}{kr_{ij}(t_0,t)}\delta\hat{\mathbf{v}}_i(t_0)\!\cdot\!\delta\hat{\mathbf{v}}_j(t_0+t)\! \!\right\rangle_{t_0}\quad\label{eq5}
\end{eqnarray}
and define the critical wavenumber $k_c=$argmax$_k\hat{C}(k,0)$, the susceptibility as $\chi=\max_k\hat{C}(k,0)$, and the correlation length as $\xi=1/k_c$ \cite{cav17,cav18,gon23}. It turns out that $k_c\propto 1/r_0$ on critical curves and we can use either the real-space or the Fourier space SCCF to find correlation length and susceptibility. 

For the DCCF, the dynamic scaling hypothesis implies 
\begin{eqnarray}
&&\frac{\hat{C}(k,t)}{\hat{C}(k,0)}= f\!\left(\frac{t}{\tau_k},k\xi\right)\!= g(k^zt,k\xi); \nonumber\\
&& g(t)=\frac{\hat{C}(k_c,t)}{\hat{C}(k_c,0)};\quad \tau_k=k^{-z}\phi(k\xi). \label{eq6}
\end{eqnarray}
Here $z$ is the dynamic critical exponent and the correlation time $\tau_k=k^{-z}\phi(k\xi)$ of the normalized DCCF (NDCCF) \eqref{eq6} at wavenumber $k$ obtained by solving the equation: \cite{cav17,gon23}
\begin{eqnarray}
\sum_{t=0}^{t_{max}} \frac{1}{t}\,\sin\!\left(\frac{t}{\tau_k}\right) f\!\left(\frac{t}{\tau_k},k\xi\right)\! = \frac{\pi}{4}.  \label{eq7}
\end{eqnarray}

An alternative definition of susceptibility uses linear response theory \cite{gon23mf}. Adding an external field, Eq.~\eqref{eq1} becomes 
\begin{widetext}
\begin{eqnarray}
\mathbf{x}_i(t+1)=\mathbf{x}_i(t)+ \mathbf{v}_i(t+1),\quad \mathbf{v}_i(t+1)=v_0  \mathcal{R}_\eta\!\left[\Theta\!\left(\sum_{|\mathbf{x}_j-\mathbf{x}_i|<R_0}\mathbf{v}_j(t)+\mathbf{H}-\beta\mathbf{x}_i(t)\right)\!\right]\!. \label{eq8}
\end{eqnarray}
We now define the vectors $\mathbf{\hat{X}}=(\mathbf{x}_1,\ldots,\mathbf{x}_N)$, $\mathbf{\hat{X}}_\alpha=((\mathbf{x}_1)_\alpha,\ldots,(\mathbf{x}_N)_\alpha)$, $\alpha=1,2,3$, $\mathbf{\hat{V}}=(\mathbf{v}_1,\ldots,\mathbf{v}_N)$ and so on. Differentiating the first equation in \eqref{eq8}, we obtain 
\begin{subequations}\label{eq9}
\begin{eqnarray} 
\mathbb{Y}_{t+1}= \mathbb{Y}_t+ \mathbb{W}_{t+1},\quad \mbox{where}\quad\mathbf{\mathcal{H}}^{\alpha\beta}=\left.\!\left(\begin{array}{c}
\frac{\partial\mathbf{\hat{X}}_\alpha}{\partial H_\beta}\\ 
\frac{\partial\mathbf{\hat{V}}_\alpha}{\partial H_\beta}\end{array}\right)\right|_{\mathbf{H}=\mathbf{0}}\! =  \left(\begin{array}{c} \mathbb{Y}\\
\mathbb{W} \end{array}\right)\!, \,\, (\mathbf{Y}^\alpha )_\beta=\mathbb{Y}^{\alpha\beta}, \,\, (\mathbf{W}^\alpha )_\beta=\mathbb{W}^{\alpha\beta}, \,\, (\bm{\delta}^\alpha)_\beta= \delta_{\alpha\beta},\quad \label{eq9a}
\end{eqnarray}
and from the second equation in Eq.~\eqref{eq8},
\begin{eqnarray} 
&&(\mathbf{W}_{t+1}^\alpha)_\beta=\left(\mathcal{R}_\eta\!\left(\mathbb{A}_{1,t}^{\alpha\gamma}\!\left[(\sigma_{R}\mathbf{w}_{1,t}^\alpha)_\gamma -\beta(\mathbf{y}_{1,t}^\alpha)_\gamma\right]\! + \mathbb{A}_{1,t}^{\alpha\gamma}(\bm{\delta}^\alpha)_\gamma\right)\!,\ldots,\mathcal{R}_\eta\!\left(\mathbb{A}_{N,t}^{\alpha\gamma}\!\left[(\sigma_{R}\mathbf{w}_{N,t}^\alpha)_\gamma -\beta(\mathbf{y}_{N,t}^\alpha)_\gamma\right]\! + \mathbb{A}_{N,t}^{\alpha\gamma}(\bm{\delta}^\alpha)_\gamma\right)\right)\!, \quad\quad \label{eq9b}\\
&&\mathbb{A}_{j,t}^{\alpha\beta}=\delta_{\alpha\beta} - \frac{[(\sigma_R\mathbf{v}_j(t))_\alpha-\beta(\mathbf{x}_j(t))_\alpha]\, [(\sigma_R\mathbf{v}_j(t))_\beta-\beta(\mathbf{x}_j(t))_\beta]}{| \sigma_R\mathbf{v}_j(t)-\beta\mathbf{x}_j|(t)^2},\quad \sigma_R\mathbf{v}_j(t)=\sum_{|\mathbf{v}_k(t)-\mathbf{v}_j(t)|<R_0}\mathbf{v}_k(t). \label{eq9c}
\end{eqnarray}\end{subequations}\end{widetext}
Here sum over repeated indices is understood. To get the last equation, we have used
\begin{eqnarray}
\delta\!\left(\frac{\mathbf{A}}{|\mathbf{A}|}\right)=\left(\mathbb{I}-\frac{\mathbf{A}\mathbf{A}^T}{|\mathbf{A}|^2}\right)\!\cdot \frac{\delta\mathbf{A}}{|\mathbf{A}|}.   \label{eq10}
\end{eqnarray}
The norm of the response matrix at zero field yields the linear response susceptibility
\begin{eqnarray}
\chi=\langle\lVert\mathbf{\mathcal{H}}_t\rVert\rangle_t, \quad \lVert\mathbf{\mathcal{H}}_t\rVert =\sqrt{\lambda_M(\mathbf{\mathcal{H}}_t\mathbf{\mathcal{H}}_t^T)}, \label{eq11}
\end{eqnarray}
where $\lambda_M(\mathbf{\mathcal{H}}_t\mathbf{\mathcal{H}}_t^T)$ is the maximum eigenvalue of the symmetric positive matrix $\mathbf{\mathcal{H}}_t\mathbf{\mathcal{H}}_t^T$ and $\langle\ldots\rangle_t$ is a time average. We find the same results replacing $\mathbb{Y}_t$ instead of $\mathbf{\mathcal{H}}_t$ in Eq.~\eqref{eq11}.

\subsection{Deterministic case $\eta=0$} 
Figure \ref{fig3} displays power laws for correlation length  $\xi$ (given either as $\xi=r_0$ where $Q(r)$ is maximum, or as the maximum swarm size), susceptibility $\chi$, winding number $w$, and correlation time $\tau$ in the limit as $\beta\to 0$.  The resulting critical exponents are relatively close to MF values and to values from numerical simulations in the noisy chaos region for larger values of $N$ \cite{gon23}.
\begin{widetext}
\begin{center}
\begin{figure}[ht]
\begin{center}
\includegraphics[clip,width=4.3cm]{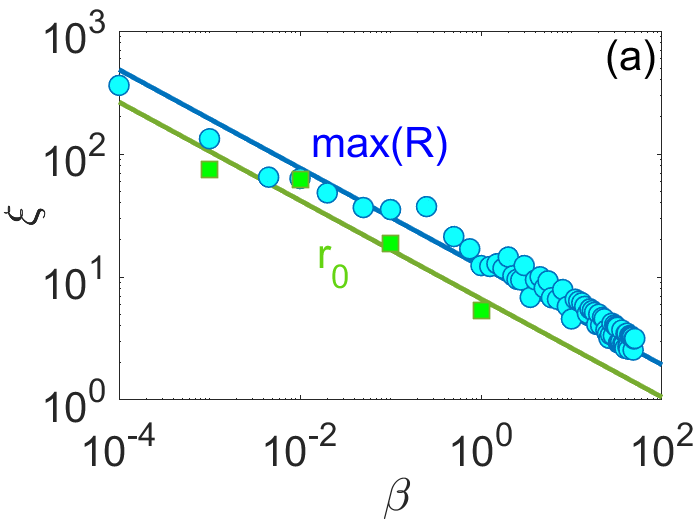}
\includegraphics[clip,width=4.3cm]{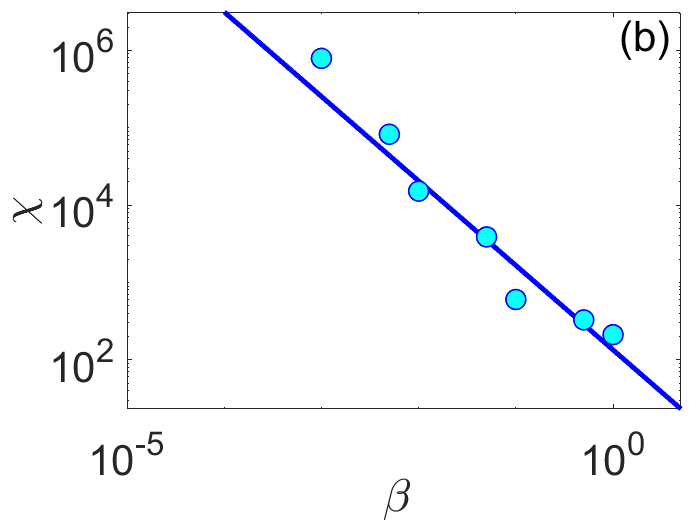}
\includegraphics[clip,width=4.3cm]{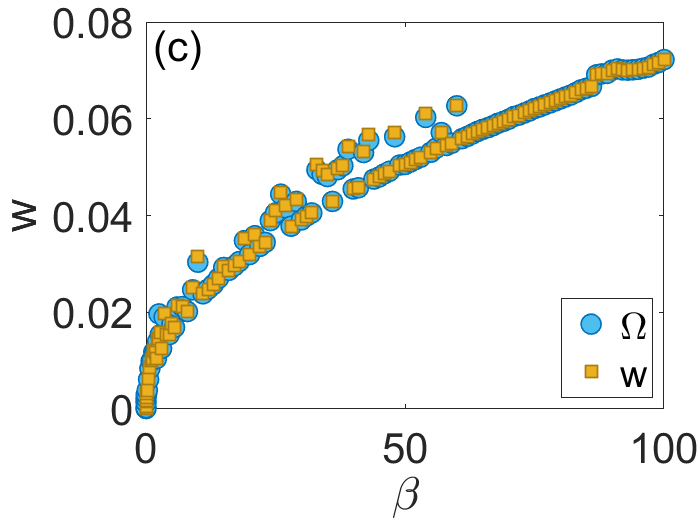}
\includegraphics[clip,width=4.3cm]{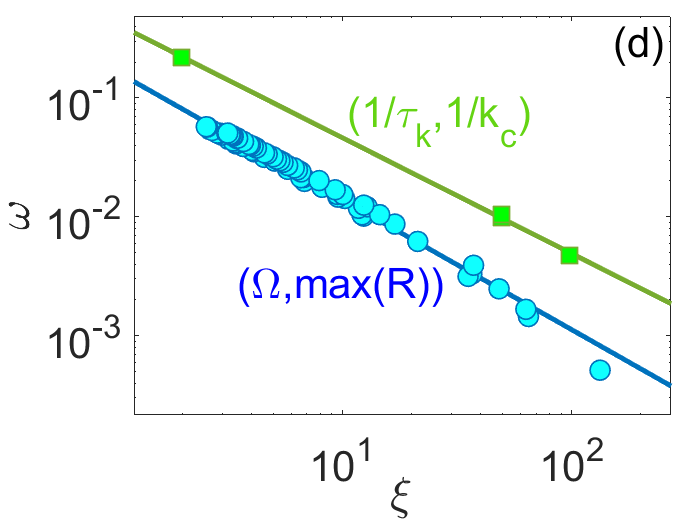}\\
\end{center}
\caption{Deterministic power laws for $N=500$ and $\eta=0$ with $\xi=r_0$ and $\xi=$ max$(R)$ vs $\beta$. {\bf (a)} $\xi\sim \beta^{-\nu}$  yield $\nu=0.40\pm0.05$. {\bf (b)} Susceptibility $\chi\sim\beta^{-\gamma}$ calculated from linear response:  $\gamma=1.09\pm 0.14$. {\bf (c)} Power law for the order parameter given by the winding number $w$ that equals the frequency $\Omega$ of the maximum spectral function \cite{gon23mf}: $\Omega=w\sim\beta^b$ with $b=0.43\pm0.01$. {\bf (d)} $\tau\sim\xi^z$: $z=1.08\pm0.06$ for $\tau=1/w$; $z=0.98\pm0.04$ for Eq.~\eqref{eq6}. The mean of these values is $\overline{z}=1.03\pm0.06$. MF values are $\gamma=z=1$, $\nu=b=\varphi=0.5$ \cite{gon23mf}.}
 \label{fig3}
\end{figure}
\end{center}
\end{widetext}

\subsection{Limit as $\eta\to 0$} 
There are three scale-free curves for which correlation length is proportional to the swarm size. For fixed $N$ and $\eta$,  consider the smallest time $t_m(\beta,N)$ at which $\hat{C}(k_c,t)=0$. $t_m(\beta,N)$ increases abruptly for a certain value $\beta_c(N;\eta)$ at which $\tau_{k_c}$ is minimum \cite{gon23}. Thus, {\em the first critical curve} $\beta=\beta_c(N;\eta)$ marks the largest possible correlation time based on the extension of $t_m(\beta,N)$ for $\beta\leq\beta_c$. As $N\to\infty$, $t_m(\beta,N)$ and $\tau_{k_c}$ tend to infinity (critical slowing down) and $\beta_c(N,\eta) \to 0$. The susceptibility and the correlation length in Eq.~\eqref{eq4} depend on $N$, $\beta$ and $\eta$. For $\beta=\beta_c$, we have the power laws:
\begin{eqnarray}
\xi\sim\beta^{-\nu},\quad \chi\sim\beta^{-\gamma}, \label{eq12}
\end{eqnarray}
as $N\to\infty$ for fixed $\eta$. Here $\nu$ and $\gamma$ are static critical exponents \cite{gon23}.  
\begin{widetext}
\begin{center}
\begin{figure}[ht]
\begin{center}
\includegraphics[trim={0.2cm 0.1cm 0.5cm 0.5cm},clip,width=4.2cm]{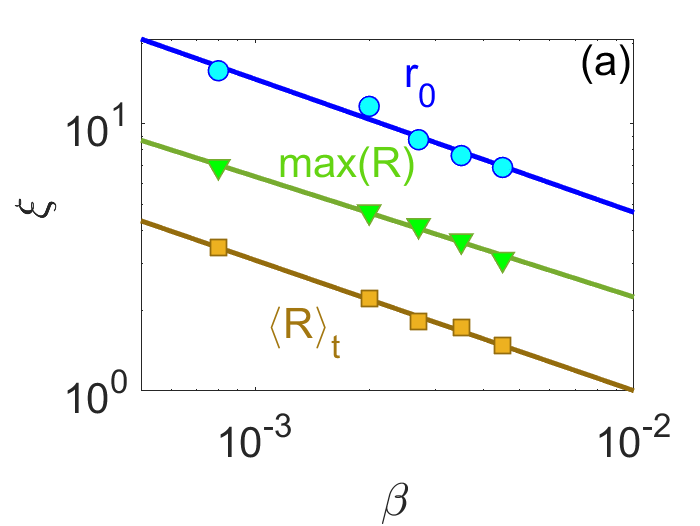}
\includegraphics[clip,width=4.2cm]{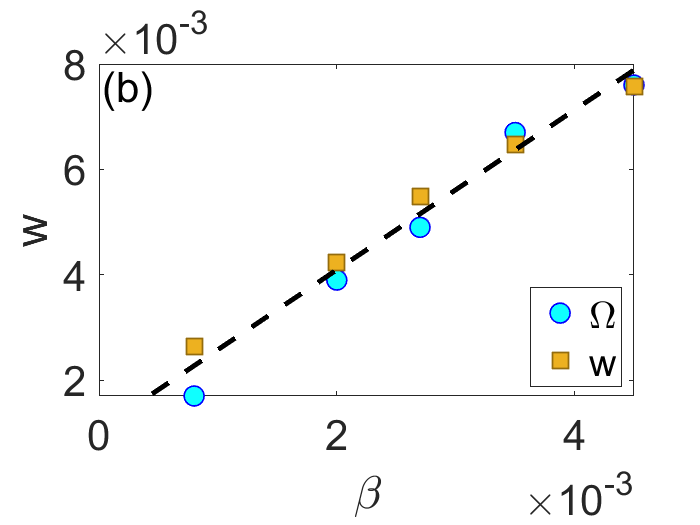}
\includegraphics[trim={0.2cm 0.1cm 0.5cm 0.5cm},clip,width=4.2cm]{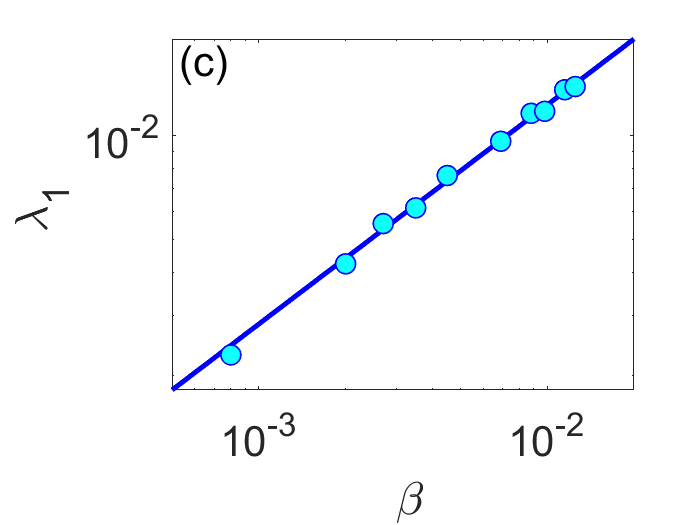}
\includegraphics[trim={0.2cm 0.1cm 0.5cm 0.5cm},clip,width=4.2cm]{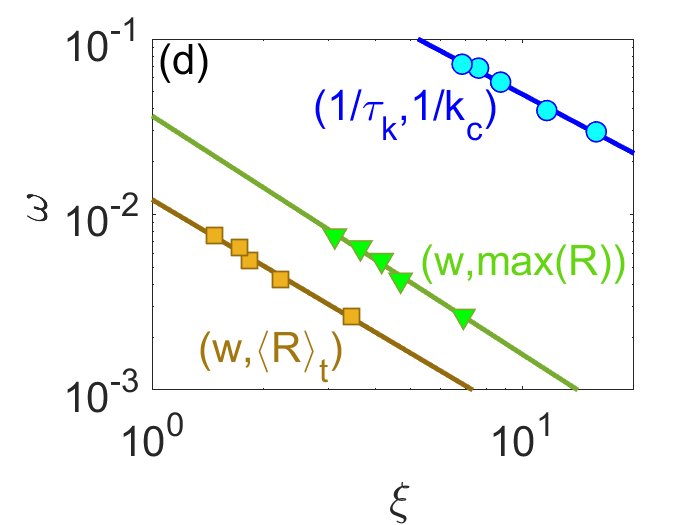}
\\
\end{center}
\caption{Critical exponents on the critical curve $\beta_c(N;\eta)$ as $(\eta,\beta_c)\rightarrow(0,0)$ for $N=500$. {\bf (a)} $\xi\sim \beta_c^{-\nu}$: $\nu=0.45\pm0.02$ for $\xi=\mbox{max}(R)$, $\nu=0.48\pm 0.05$ for $\xi=\langle R \rangle_t$, and  $\nu=0.48\pm 0.06$ for $\xi=r_0$. The mean of these values is $\overline{\nu}=0.47\pm0.06$. {\bf (b)} Winding number vs $\beta_c$: $w\sim\beta_c^b$, $b=0.60\pm0.03\approx z\nu$ ($b=0.5$ for the MF theory \cite{gon23mf}).  {\bf (c)} LLE vs $\beta_c$: $\lambda_1\sim\beta_c^\varphi$, $\varphi=0.62\pm0.06\approx z\nu$. {\bf (d)} Dynamical critical exponent $z$, $\tau\sim\xi^z$: $z=1.35\pm0.09$ for $\tau=1/w$ ($w$ is the winding number \cite{gon23mf}) and $\xi=\mbox{max}(R)$, $z=1.25\pm 0.08$ for $\tau=1/w$, $\xi=\langle R \rangle_t$, and $z=1.12\pm0.07$ for Eq.~\eqref{eq6}. Mean: $\overline{z}= 1.24\pm 0.08$. }
 \label{fig4}
\end{figure}
\end{center}

\begin{center}
\begin{figure}[ht]
\begin{center}
\includegraphics[clip,width=4.3cm]{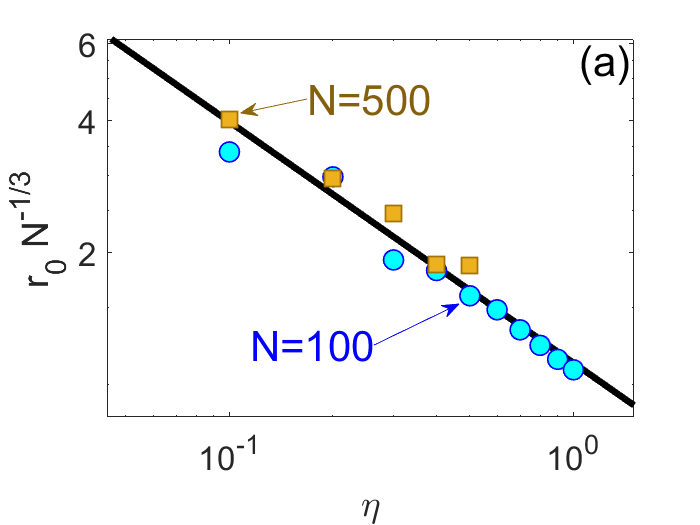}
\includegraphics[clip,width=4.3cm]{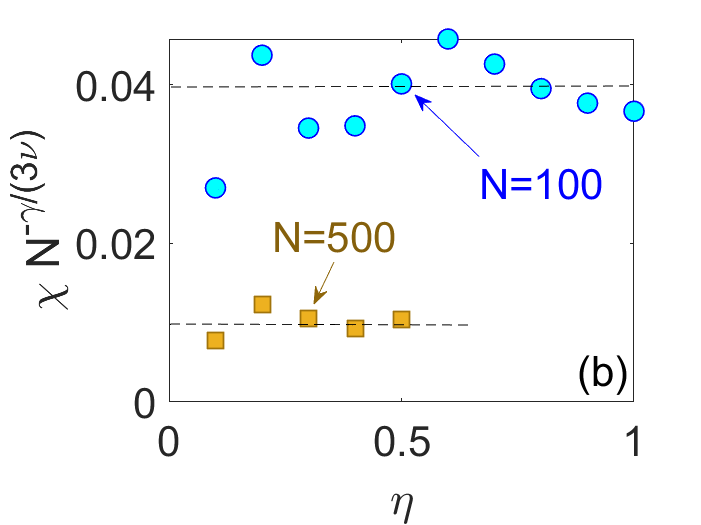}
\includegraphics[clip,width=4.3cm]{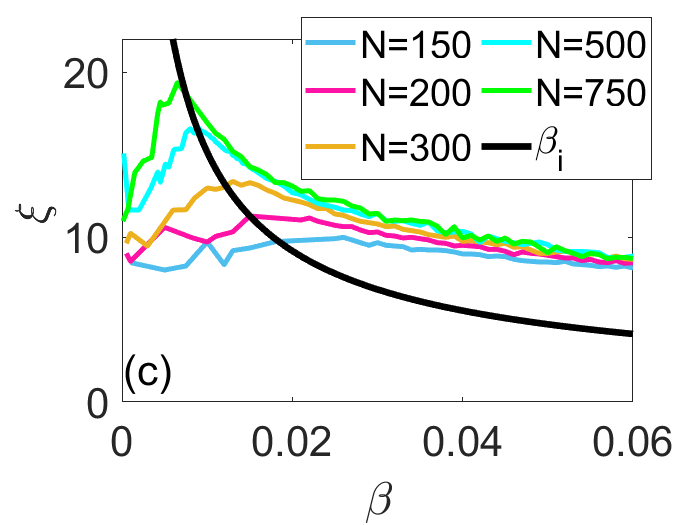}
\includegraphics[clip,width=4.3cm]{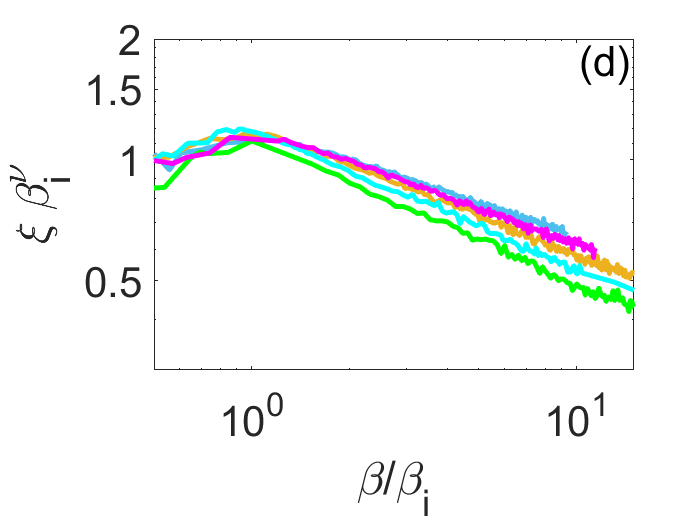}
\\
\end{center}
\caption{ {\bf (a)} Scaled correlation length $\xi N^{-\frac{1}{3}}$ and  {\bf (b)} susceptibility $\chi N^{-\frac{\gamma}{3\nu}}$ vs noise for $\beta_c(N;\eta)$, $N=100,500$.  {\bf (c)} For $\eta=0.5$, $\xi$ vs $\beta$; the black curve marks the local maxima and fits $\xi \sim\beta_i^{-\nu}$. {\bf (d)} Rescaled curves $\xi\beta_i^\nu$ vs $\beta/\beta_i$ show collapse of the curves in (c) to a plateau.}
 \label{fig5}
\end{figure}
\end{center}

\begin{center}
\begin{figure}[ht]
\begin{center}
\includegraphics[clip,width=4.3cm]{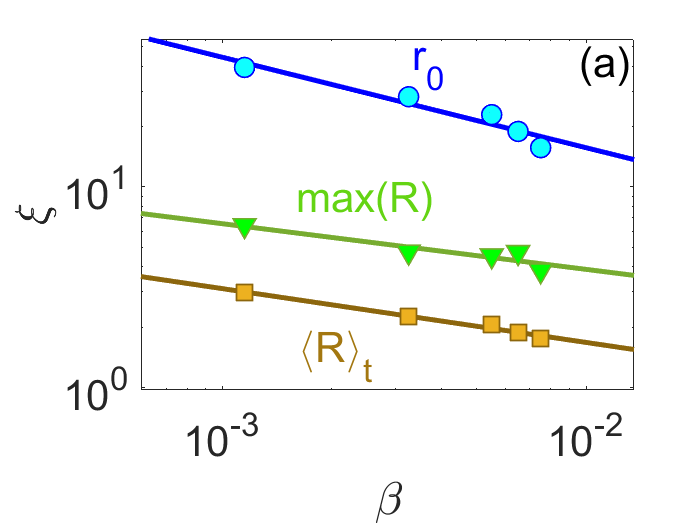}
\includegraphics[clip,width=4.3cm]{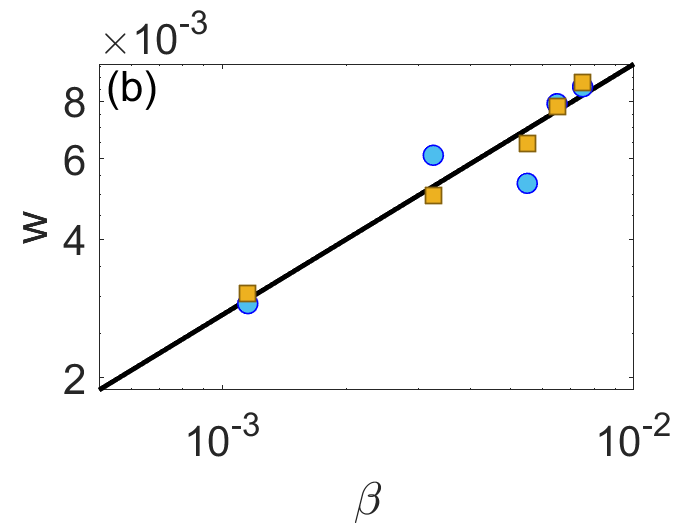}
\includegraphics[clip,width=4.3cm]{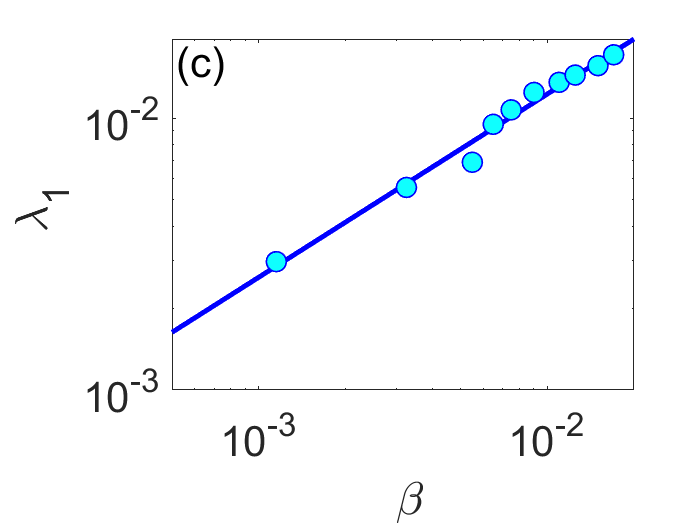}
\includegraphics[clip,width=4.3cm]{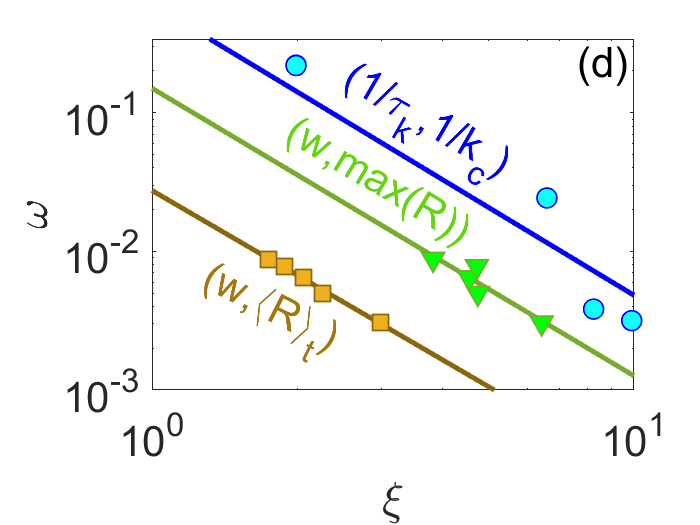}
\\
\end{center}
\caption{Same as Figure \ref{fig4} for the critical curve $\beta_i(N;\eta)$. {\bf (a)} $\xi\sim \beta_i^{-\nu}$: $\nu=0.23\pm0.05$ for  $\xi=\mbox{max}(R)$, $\nu=0.27\pm 0.07$ for $\xi=\langle R \rangle_t$, $\nu=0.44\pm0.06$ for $\xi=r_0$. The latter value is close to that for $\beta_c(N;\eta)$, but not the other values. {\bf (b)} Winding number vs $\beta_i$: $b=0.55\pm0.05$. {\bf (c)} LLE exponent: $\varphi=0.68\pm0.06\approx z\nu$. 
{\bf (d)} $z=2.07\pm0.30$, $z=2.02\pm0.09$ and $z=2.10\pm0.11$ from $(w,\mbox{max}(R))$, $(w,\langle R \rangle_t)$ and $(\tau_k, \xi=1/k_c)$, respectively. Mean of values: $\overline{z}=2.06\pm0.14$.}
 \label{fig6}
\end{figure}
\end{center}

\begin{center}
\begin{figure}[ht]
\begin{center}
\includegraphics[clip,width=4.3cm]{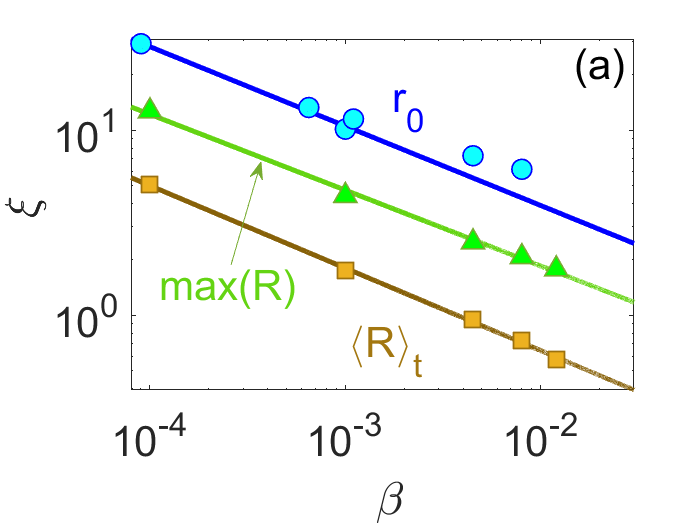}
\includegraphics[clip,width=4.3cm]{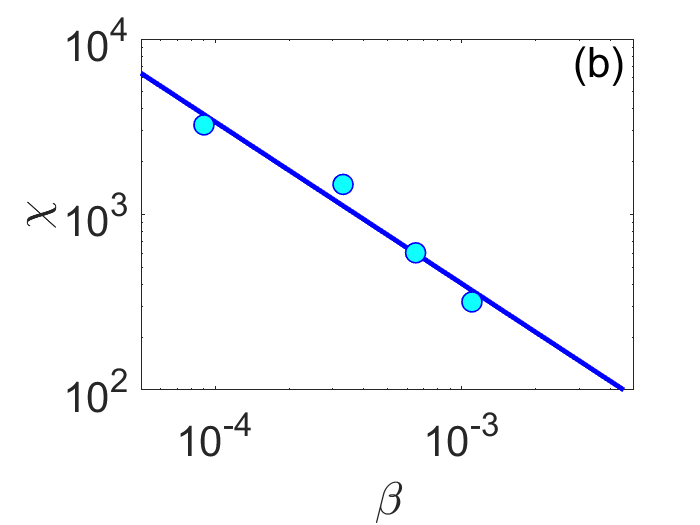}
\includegraphics[clip,width=4.3cm]{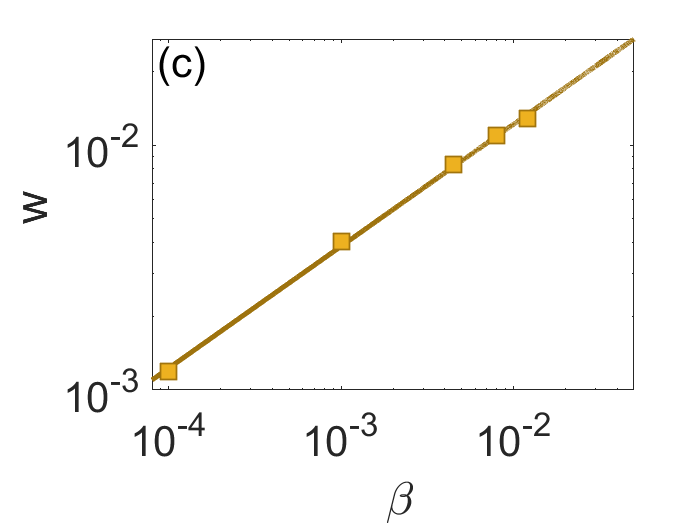}
\includegraphics[clip,width=4.3cm]{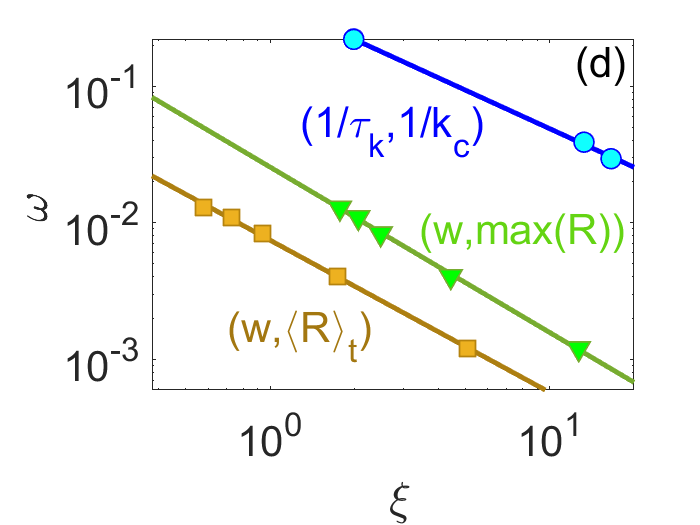}\\
\end{center}
\caption{Critical exponents on the critical curve $\beta_0(N;\eta)$ as $(\eta,\beta_0)\rightarrow(0,0)$ for $N=500$. {\bf (a)} $\nu=0.41 \pm 0.02$, $\nu=0.44\pm 0.01$ and $\nu=0.43\pm0.03$ for $\xi=\mbox{max}(R)$, $\xi=\langle R \rangle_t$ and $\xi=r_0$, respectively. Mean of values: $\overline{\nu}=0.43\pm0.03$. {\bf (b)} Linear response susceptibility $\chi\sim\beta^{-\gamma}$ with $\gamma=0.92\pm0.13$. {\bf (c)} Winding number vs $\beta_0$: $b=0.50\pm0.01\approx z\nu$. {\bf (d)} $z=1.20\pm0.06$, $z=1.12\pm0.04$ and $z=1.06\pm0.08$ from $(w,\mbox{max}(R))$, $(w,\langle R \rangle_t)$ and $(\tau_k, \xi=1/k_c)$, respectively. Mean of values: $\overline{z}=1.12\pm0.08$.}
 \label{fig7}
\end{figure}
\end{center}\end{widetext}

Including the dependence on noise, we have found Eq.~\eqref{eq2} and
\begin{subequations}\label{eq13}
\begin{eqnarray}
&& r_{0j}=D_j \beta_j^{-\nu} \eta^{-p_j}=\frac{D_j }{C_j^\nu}N^\frac{1}{3} \eta^{-p_j - \nu m_j}, \quad j=0,c,\quad \label{eq13a}\\
&&\chi = Q(r_0)=  Q_j\beta_j^{-\gamma}\eta^{q_j} =  \frac{Q_j}{C_j^\gamma}N^\frac{\gamma}{3\nu}\eta^{q_j-\gamma m_j}.     \label{eq13b}
\end{eqnarray}\end{subequations}
Fig.~\ref{fig4}(a) shows that $\xi$ measured with $r_0$, the maximum value of the center of mass length, $R=|\mathbf{X}|$, or its time average, $\langle R\rangle_t$,  scale as $\beta_c(N;\eta)^{-\nu}$ as $\eta\to 0$, $\beta_c\to 0$. Similarly, the power laws for winding number and LLE vs $\beta$ are displayed in Figs.~\ref{fig4}(b) and \ref{fig4}(c), respectively, where  Fig.~\ref{fig4}(d) produces the dynamic critical exponent $z$ using different definitions. The values of the critical exponents are similar to those in Fig.~\ref{fig3}. There is little dispersion in the exponent $\nu$, but the dispersion is larger for $z$. Using data from $N=100,500$ and Figs.~\ref{fig5}(a) and \ref{fig5}(b), we have obtained the parameters $C_c=1.5\pm 0.2$, $D_c= 1.33 \pm 0.04$, $p_c\approx 0$ ($p_c + \nu m_c = 0.55 \pm 0.03$), $q_c\approx\gamma m_c$ in Eqs.~\eqref{eq13}. At fixed $N=500$, the critical line $\beta_c(N;\eta)$ becomes zero as $\eta\to 0$; see Fig.~\ref{fig1}(b). Then $\xi=r_0\to\infty$ as $\eta\to 0$ for fixed $N$ and there are power laws $\xi\ N^{-\frac{1}{3}}\sim\eta^{-p_c-\nu m_c}$ and $\tau\sim \xi^z$ but $\chi N^{-\frac{\gamma}{3\nu}}$ seems to be independent of noise and $Q_c/C_c^\gamma$ depends on $N$; see Fig. \ref{fig5}(b). 

The {\em second critical curve} corresponds to the inflection point of the susceptibility, $\beta_i(N;\eta)$, for fixed $N$ and $\eta$. It turns out that on this line, the largest Lyapunov exponent (LLE) reaches a local maximum \cite{gon23}. Fig.~\ref{fig5}(c) shows that the correlation length also has a local maximum at $\beta_i$. Fig.~\ref{fig5}(d) shows that $\xi\sim \beta_i^{-\nu}$, same as for the critical line $\beta_c$. The critical exponents for $\beta_i(N;\eta)$ are different from those of the other curves and from MF values, as shown in Figure \ref{fig6}.  

\begin{center}
\begin{figure}[ht]
\begin{center}
\includegraphics[clip,width=4.2cm]{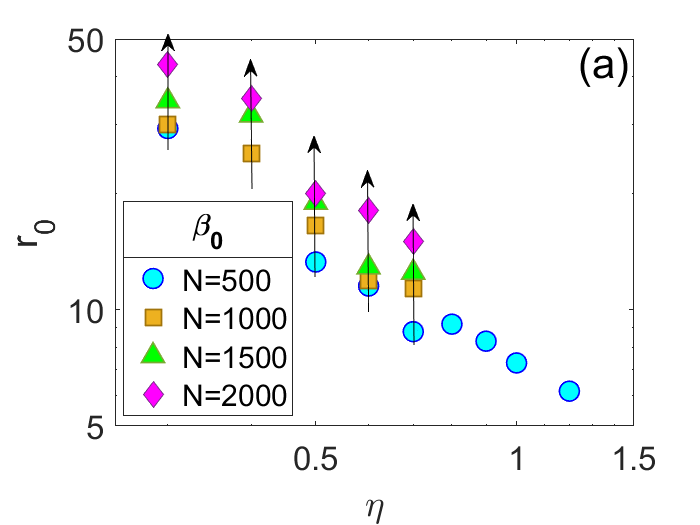}
\includegraphics[clip,width=4.2cm]{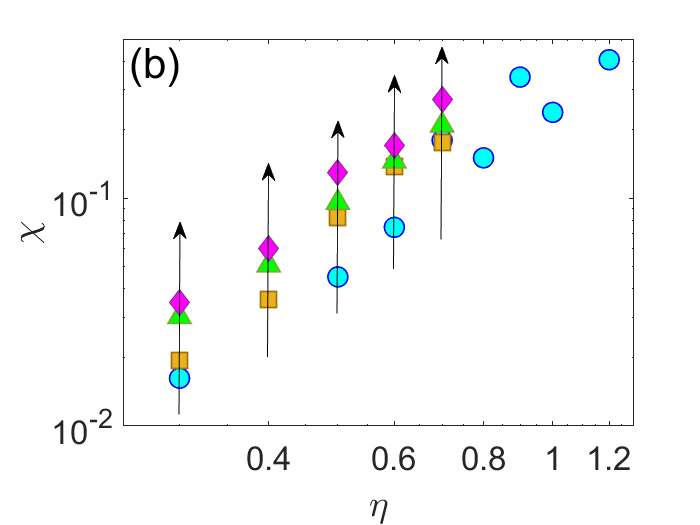}
\includegraphics[clip,width=4.2cm]{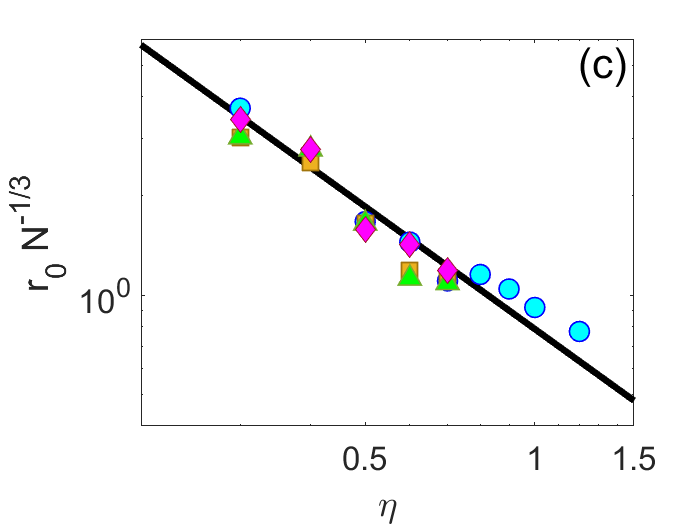}
\includegraphics[clip,width=4.2cm]{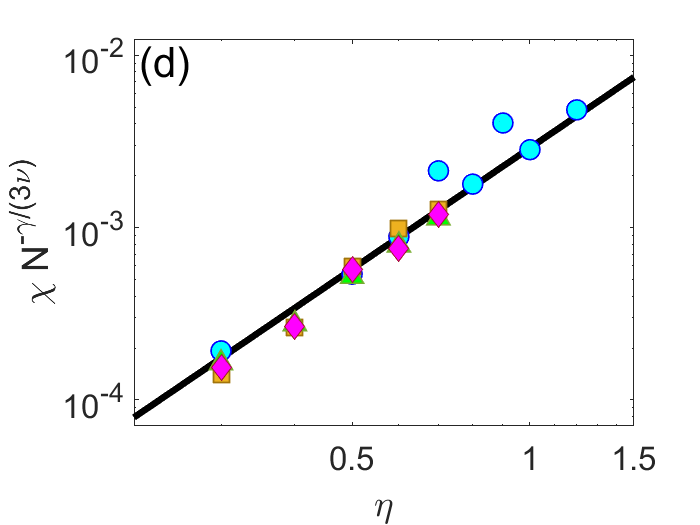}\\
\end{center}
\caption{Critical exponents on the critical curve $\beta_0(N;\eta)$ as $(\eta,\beta_0)\rightarrow(0,0)$ for $N=500, 1000, 1500, 2000$. {\bf (a)} $\nu=0.43\pm0.03$; {\bf (b)} $\gamma=0.92\pm0.13$; {\bf (c)} $\xi N^{-\frac{1}{3}}$ vs $\eta$; {\bf (d)} $\chi N^{-\frac{\gamma}{3\nu}}$ vs $\eta$. Parameter values: $D_0= 0.76 \pm 0.08$, $p_0 + \nu m_0 = 1.24 \pm 0.11$ ($p_0 = 1.24 - \nu m_c - \nu a_2 N^{-n_2} = 0.72 - \nu a_2 N^{-n_2}$). $Q_0=0.003$, $q_0-\gamma m_0=2.39\pm0.17$ ($q_0=2.39+\gamma m_c+\gamma a_2 N^{-n_2}=3.5+\gamma a_2 N^{-n_2}$).}
 \label{fig8}
\end{figure}
\end{center}

The {\em third critical curve} $\beta_0(N;\eta)$ separates the region of single cluster chaos from the non-chaotic region $0<\beta< \beta_0(N;\eta)$; see Fig.~\ref{fig1}(a) and Figure 1 of \cite{gon23mf}. While $\beta_0$ separating non-chaotic and single-cluster chaotic regions was shown to be scale-free in the MF approximation, this curve was not studied in Ref.~\cite{gon23}.  Fig.~\ref{fig7}(a) and \ref{fig7}(b) yield the critical exponents $\nu$ and $\gamma$, respectively. They are comparable to those produced in the deterministic case, the MF approximation, and those found from numerical simulations in the noisy chaos region for larger values of $N$ \cite{gon23}. Note that the linear response susceptibility produces a power law and critical exponent for the curve $\beta_0(500;\eta)$ because the algorithm defining it \cite{gon23mf} converges. Including values with $N=500, 1000, 1500, 2000$, we get Fig.~\ref{fig8} for correlation length $\xi=r_0$ and susceptibility $\chi=Q(r_0)$ versus noise.  While $\xi\to\infty$ as $\eta\to 0$ (critical exponent $\nu=0.43$), $\chi\to 0$ because $q_0-\gamma m_0=2.4\pm 0.2>0$ as shown in ~\ref{fig8}(d). Why? $\xi=r_0\to \infty$ as $\eta\to 0$ because $-p_0-\nu m_0<0$ in Eq.~\eqref{eq13a}. Then the step function in Eq.~\eqref{eq4} is always 1 and $Q(r_0)=0$ because $\sum_{j=1}^N\delta\mathbf{\hat{v}}_j=0$. On the other hand, the linear response susceptibility given by Eq.~\eqref{eq11} goes to infinity as $\beta_0\to 0$ with the correct critical exponent $\gamma=0.92$ shown in Fig.~\ref{fig7}(b).

The three critical curves do not change if we redefine the average swarm velocity in Eqs.~\eqref{eq2} and \eqref{eq13} by subtracting overall rotations and dilations from $\mathbf{V}$ at each time step \cite{gon23}. There is another scale-free line obtained by tracking the local maxima of the susceptibility as a function of $\beta$. On this line, the chaotic swarm comprises several clusters and  rotations and dilations are noticeable. However, when we subtract overall rotations and dilations, the local maxima of the susceptibility disappear \cite{gon23}, which is why we do not add this line to the previous list of three critical lines.  Note that we can find the static critical exponents $\nu$ and $\gamma$ from data at a fixed $N=500$ as in Figs.~\ref{fig7}(a) and \ref{fig7}(b) or from rescaled correlation length and susceptibility for different values of $N$ as in Figs.~\ref{fig8}(c) and \ref{fig8}(d).

\begin{center}
\begin{figure}[ht]
\begin{center}
\includegraphics[clip,width=4.2cm]{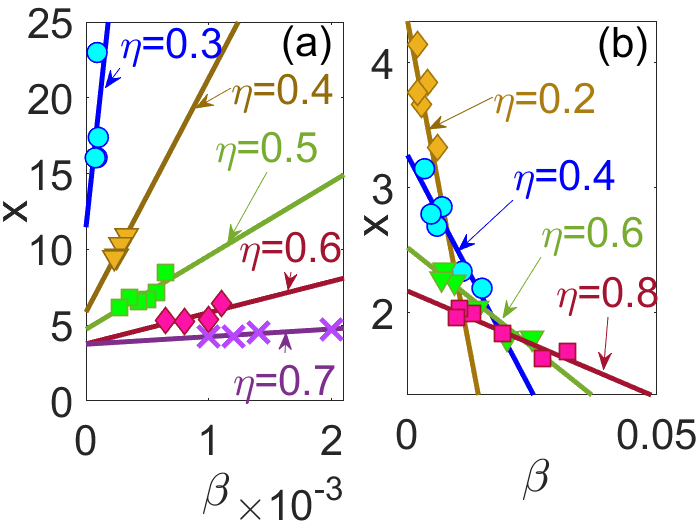}
\includegraphics[clip,width=4.2cm]{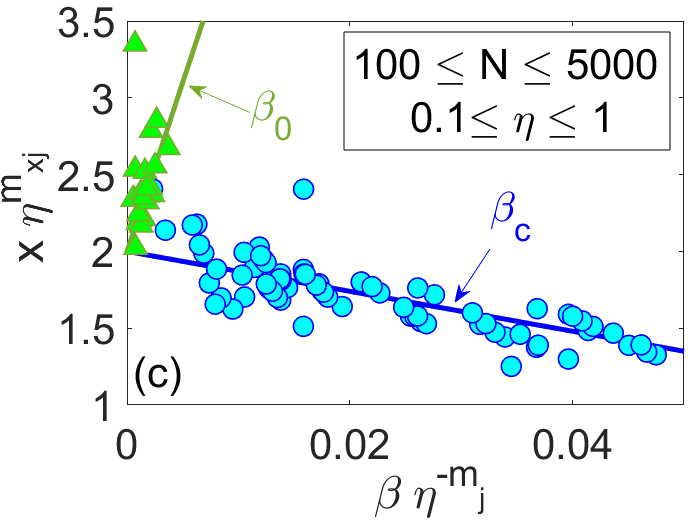}
\\
\end{center}
\caption{Perception range. {\bf (a)} $x$ vs $\beta_0$ for $100\leq N\leq 500$ and $\eta$ as marked. {\bf (b)} Same for $\beta_c$. {\bf (c)} $\eta^{m_{xj}}x$ vs $\eta^{-m_j}\beta_j$, $j=0,c$, for $100\leq N\leq 5000$ and $0.1\leq\eta\leq 1$. }
 \label{fig9}
\end{figure}
\end{center}

\subsection{Perception range and static critical exponents} 
While confinement or noise are not measurable control parameters, the perception range $x$ (time averaged arithmetic mean of the minimal distance between each particle and its closest neighbor \cite{att14,att14plos}) is. For $\beta_0(N;\eta)$ and finite $N$, $x>x_c(\eta)$ (the value at $N=\infty$), whereas $x<x_c(\eta)$ on $\beta_c(N;\eta)$; see Figs.~\ref{fig9}(a) and \ref{fig9}(b). In terms of the perception range, static critical exponents are defined by \cite{att14} 
\begin{eqnarray}
\xi\sim(x-x_c)^{-\nu},\quad \chi\sim(x-x_c)^{-\gamma},\label{eq14}
\end{eqnarray}
which are similar to Eq.~\eqref{eq12}. Fig.~\ref{fig9}(c) shows that there is a linear relation between rescaled versions of $x$ and $\beta_0$ or $\beta_c$:
 \begin{eqnarray}
\eta^{m_{x0}}x=A_0+B_0\beta_0\eta^{-m_0},\,\eta^{m_{xc}}x=A_c-B_c\beta_c\eta^{-m_c}\!.\quad\, \label{eq15}
\end{eqnarray}
Here the critical perception range at zero confinement is $x_c(\eta)=A_j\eta^{-m_{xj}}$, $j=0,c$, and we have found $m_{xc}=0.50\pm 0.03$, $A_c=2.00\pm 0.02$, $B_c=13.00\pm0.03$, $m_{x0} = 1.6\pm 0.2$, $A_0=2.0 \pm 0.2$, $B_0=219.8\pm 0.2$ from numerical simulations. Eqs.~\eqref{eq2} and \eqref{eq15} imply that $\xi\sim N^\frac{1}{3}\sim\eta^{\nu m_j}\beta_j(N;\eta)^{-\nu}\sim\eta^{\nu m_{xj}}|x-x_c|^{-\nu}$, $j=0,c$. Similarly,  $\chi\sim|x-x_c|^{-\gamma}$ from Eqs.~\eqref{eq12} and \eqref{eq15}. Thus, the empirical relations \eqref{eq14} have the same static critical exponents as the relations \eqref{eq12}. 

\section{Mixtures of simulation data in extended criticality regions}\label{sec:4} 
Natural swarms experience background noise and variable atmospheric conditions \cite{ni15epj} that may account for their strong correlations \cite{att14,cav17,cav18}. The measured power law for macroscopic quantities indicate that swarms are close to criticality. To interpret measurements using the HCVM, we need to recreate a mixture of results of numerical simulations that resemble measurements taken from swarms of different $N$, $\eta$ and $\beta$, all within the extended criticality region [III] of  Fig.~\ref{fig1}(a). With such a mixture, we find $\nu=0.43\pm 0.03$, $\gamma=0.92\pm 0.13$, close to the observed values: $\nu=0.35\pm 0.1$, $\gamma= 0.9\pm 0.2$ \cite{att14}.

\begin{widetext}
\begin{center}
\begin{figure}[ht]
\begin{center}
\includegraphics[clip,width=4.3cm]{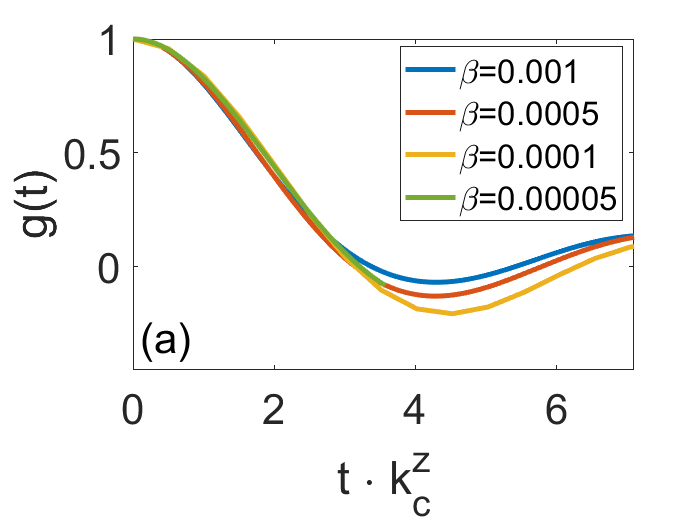}
\includegraphics[clip,width=4.3cm]{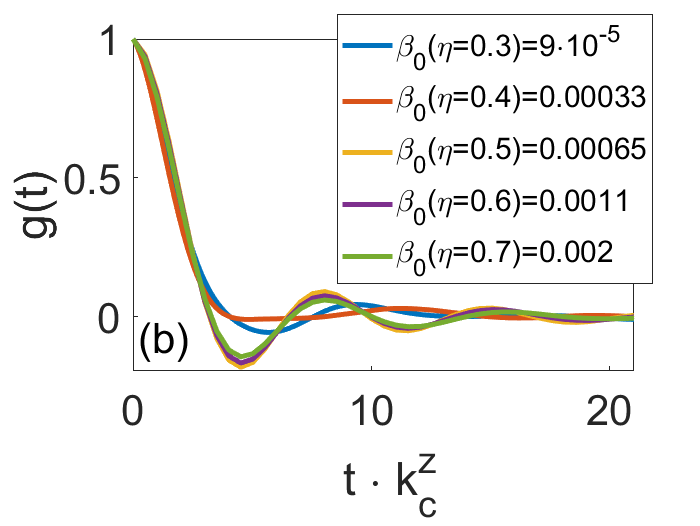}
\includegraphics[clip,width=4.3cm]{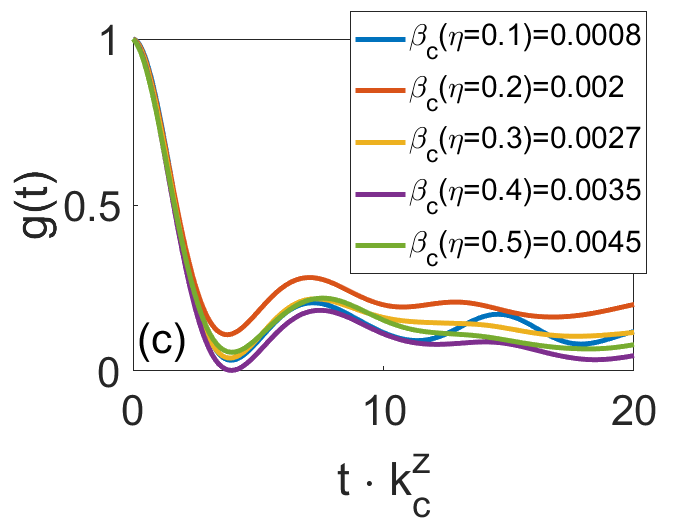}
\includegraphics[clip,width=4.3cm]{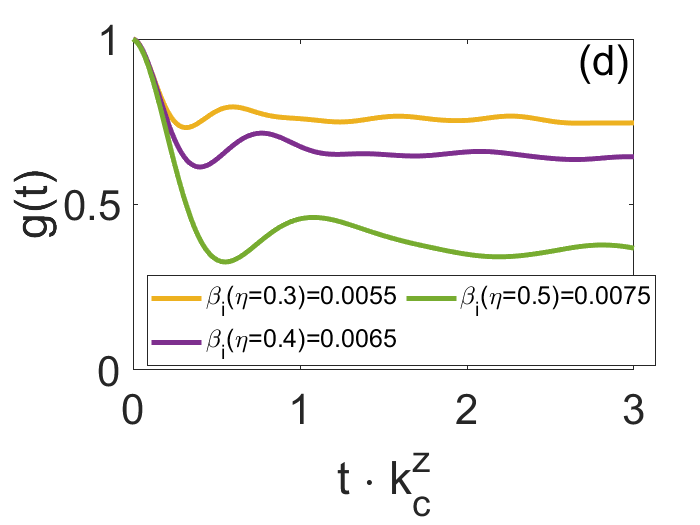}
\\
\end{center}
\caption{Collapse of the NDCCF $g(t)$ using mixtures of simulation data with $N=500$. {\bf (a)} Mixture of deterministic data ($\eta=0$) with the values of $\beta$ listed in the inset yielding $z\approx1$. {\bf (b)} Mixture of data on the critical line $\beta_0$ with the values of $\eta$ listed in the inset yielding $z\approx1$. {\bf (c)} Mixture of data on the critical line $\beta_c$  with the values of $\eta$ listed in the inset yielding $z\approx1.1$. {\bf (d)} Mixture of data on the critical line $\beta_i$ with the values of $\eta$ listed in the inset yielding $z\approx2.1$.}
 \label{fig10}
\end{figure}
\end{center}\end{widetext}

\subsection{Collapse of NDCCF data and exponent $z$}
We will use a mixture of data from the critical lines $\beta_0(N;\eta)$ and $\beta_c(N;\eta)$ but not from the line of maximal LLE, $\beta_i(N;\eta)$. Why? Firstly, at $\beta_i(N;\eta)$, the swarm starts developing several clusters whereas it comprises a single cluster on the other critical lines. Secondly, the intervals of scaled times $k_c^zt=t\xi^{-z}$ over which NDCCF data collapse are similar for $\beta_0(N;\eta)$ and $\beta_c(N;\eta)$ but it is much smaller for $\beta_i(N;\eta)$. Here the correlation length $\xi$ is $1/k_c$, where $k_c=$argmax$_k\hat{C}(k,0)$ for the Fourier transform of the DCCF in Eq.~\eqref{eq5} calculated on the critical curve \cite{gon23}. Figure \ref{fig10} illustrates the collapse of NDCCF data for values on the deterministic line $\eta=0$ as $\beta\to 0$ and for values on the three critical curves as $\eta\to 0$. In all cases, NDCCF data collapse only for short scaled times $k_c^z t$ on intervals $(0,\delta)$, where $\delta$ ranges from 0.25 to 4. The critical dynamical exponent $z$ is near 1 for $\beta_0$ and $\beta_c$ and the interval width $\delta$ is similar for these curves. However, the critical curve $\beta_i(N;\eta)$ has $z\approx 2.1$ and the smallest value of $\delta$, which indicates different dynamics and may correspond to different length scales in the multifractal chaotic attractors associated to $\beta_i(N;\eta)$ \cite{gon23}. The LLE is greatest on $\beta_i$. Moreover, the swarm may start splitting into different clusters for this stronger confinement, but it is formed by a single cluster for $\beta_0$ and $\beta_c$. As the swarm shape, $z$ value and width $\delta$ are similar for curves $\beta_0$ and $\beta_c$, we select values for mixtures of data only on these curves of the extended critical region. 
\begin{center}
\begin{figure}[ht]
\begin{center}
\includegraphics[clip,width=4.2cm]{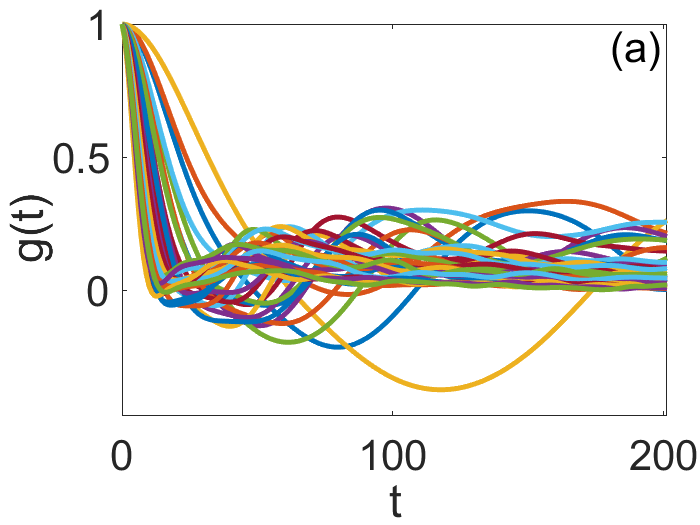}
\includegraphics[clip,width=4.2cm]{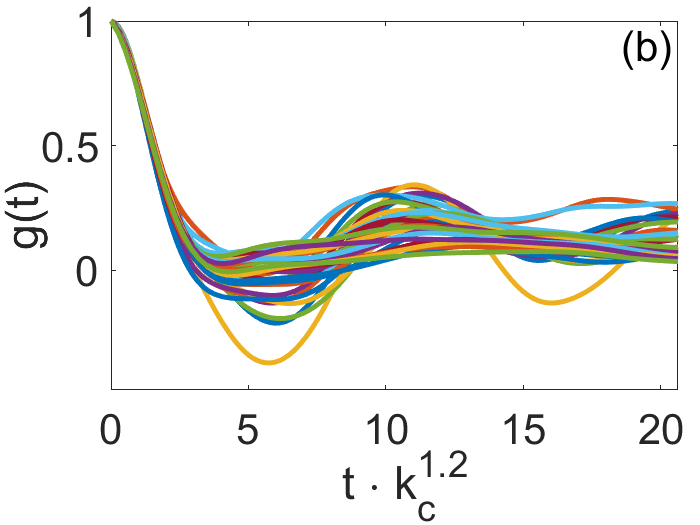}
\\
\end{center}
\caption{Normalized dynamic connected correlation function (NDCCF) $g(t)=\hat{C}(k_c,t)/\hat{C}(k_c,0)$ with $k_c\!=$ argmax$_k\hat{C}(k_c,0)=1/\xi$ \cite{gon23} for $\beta=\beta_c(N;\eta)$, $0.1\leq\eta\leq1$ and $100\leq N\leq 300$. {\bf (a)} $g(t)$. {\bf (b)} Visual collapse of the NDCCF as a function of $k_c^zt$ for $z\approx1.2$; $z_\text{LS}=z_\text{RMA}=1.09\pm0.02$. } 
 \label{fig11}
\end{figure}
\end{center}

The observed dynamical critical exponents are $z=1.2$ \cite{cav17}, and, with more data points, $z_\text{LS}=1.16\pm 0.12$ \cite{cav23} (Supplementary Material, calculated using LS regression; see below). Using HCVM numerical simulation data with $\eta=0.5$ and particle numbers $100\leq N\leq5000$ on $\beta_c(N;\eta)$, the NDCCF occurs on the same interval of scaled times (Fig.~4 of \cite{gon23}), and $z= 1.01\pm 0.01$. Fixing $N=500$, a mixture of HCVM data on $\beta_0(N;\eta)$ for noises between 0.3 and 0.7 yields $z\approx 1$, whereas a similar mixture for noises between 0.1 and 0.5 on $\beta_c$ gives $z\approx 1.1$; see Fig.~\ref{fig10}. A wider noise interval $0.1\leq\eta\leq 1$ on the critical curve $\beta_c(N;\eta)$  for $100\leq N\leq 300$ produces the data collapse shown in Fig.~\ref{fig11}, and the exponent $z=1.09\pm0.02$. 

\subsection{Dynamical critical exponent $z$} 
We use data extracted from numerical simulations of the HCVM for the critical curves $\beta_0$ and $\beta_c$ for different noise values and particle numbers. Figure \ref{fig12} shows the results of using a variety of data from numerical simulations of the HCVM to determine $z$. Figure \ref{fig12}(a) depicts correlation time vs correlation length for points on the scale free curves $\beta_0$ and $\beta_c$. Using LS regression, $z\approx 1$ for $\beta_0$ and $\beta_c$. The standard deviation $\sigma$ is larger for $0.1\leq \eta\leq 0.5$ than for larger $\eta$ and so is the difference $\Delta\beta=\beta_c-\beta_0$; see Fig.~\ref{fig12}(b). Natural swarms have relatively small sizes (the largest observed swarm has $N=781$) and data are inevitably noisy \cite{cav17,cav23}. Thus, we select data points on scale-free curves $\beta_0(N;\eta)$ and $\beta_c(N;\eta)$ for $0.1\leq \eta\leq 0.5$ (critical region [III] in Fig.~\ref{fig1}) to calculate $z$ in Fig.~\ref{fig12}(c). 

As explained in \cite{cav23}, fitting a straight line by RMA regression takes into consideration both the errors in $\tau$ and $\xi$, whereas LS regression considers only the error in $\tau$, thereby underestimating $z$. If we take data points on a single critical line, as in Figures \ref{fig10} and \ref{fig11}, the values of the dynamic critical exponent are the same whether we calculate $z$ using LS or RMA regression. However, for a mixture of data on the lines $\beta_0$ and $\beta_c$ with $0.1\leq\eta\leq 0.5$ and $100\leq N\leq 2500$, we find  $z_\text{LS}=1.15\pm 0.11$ and $z_\text{RMA}=1.33\pm 0.10$ with probability distributions shown in Fig.~\ref{fig12}(d). A different mixture of data with $N=500$ and $0.1\leq\eta\leq 0.5$ yields the values $z_\text{LS}=1.24\pm 0.11$ and $z_\text{RMA}=1.37\pm 0.10$, shown in Figure \ref{fig13}.

All these values are very close to measurements on natural swarms: $z_\text{LS}=1.16\pm 0.12$ and $z_\text{RMA}=1.37\pm 0.11$ \cite{cav23}. They are also close to the RG prediction $z=1.35$ for the ordering transition of the active version of models E/F and G in Ref.~\cite{hoh77}. Numerical simulations of the ordering transition of the inertial spin model (ISM \cite{cav15}) with periodic boundary conditions yield $z_\text{LS}=z_\text{RMA}=1.35\pm 0.04$ \cite{cav23}. 

\begin{widetext}
\begin{center}
\begin{figure}[ht]
\begin{center}
\includegraphics[clip,width=16cm]{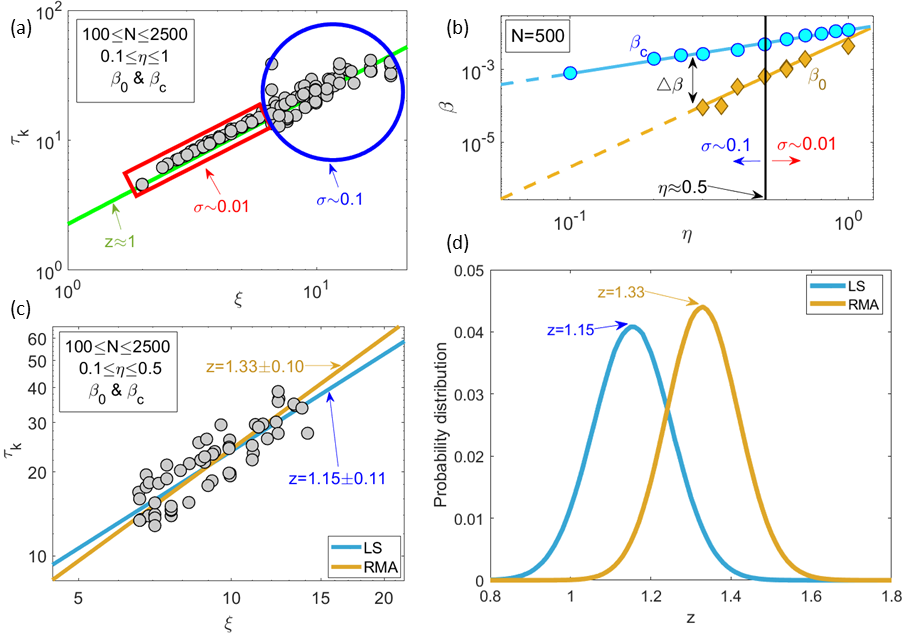}
\\
\end{center}
\caption{Mixtures of simulation data. {\bf (a)} Correlation time vs length for data on scale-free curves $\beta_0(N;\eta)$ and $\beta_c(N;\eta)$ for $0.1<\eta<1$ and $100<N<2500$. Note that the standard deviation $\sigma$ is larger for smaller noise values. {\bf (b)} Scale-free curves for $N=500$ showing the noise intervals with smaller and larger $\sigma$. $\Delta\beta=\beta_c-\beta_0$ increases as $\eta$ decreases.  {\bf (c)} Same as panel (a) for $0.1<\eta<0.5$ showing LS and RMA fittings to straight lines for $\beta_0$ and $\beta_c$ data: the corresponding dynamical critical exponents are $z_\text{LS}=1.15\pm 0.11$ and $z_\text{RMA}=1.33\pm 0.10$.   {\bf (d)} Probability distribution of the LS (blue) and RMA (orange) critical exponent $z$ from the resampling method consisting of randomly drawing $10^7$ subsets with half the number of points from numerical simulations. Then we determine $z$ in each subset using LS and RMA \cite{cav23}. \label{fig12}}
\end{figure}
\end{center}

\begin{center}
\begin{figure}[ht]
\begin{center}
\includegraphics[clip,width=6.8cm]{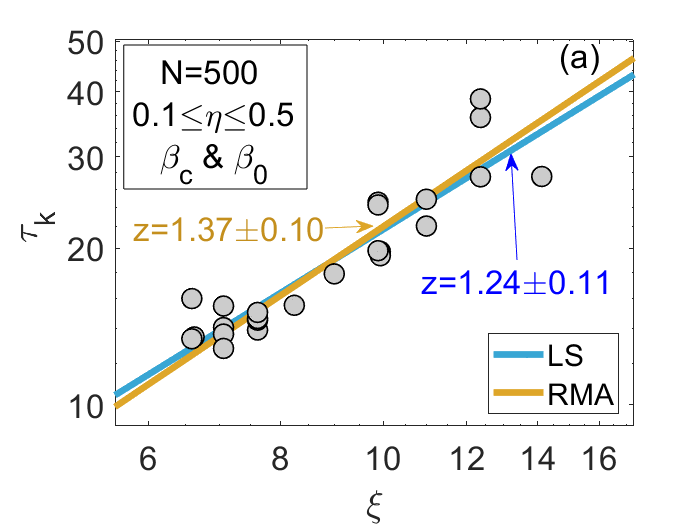}
\includegraphics[clip,width=6.8cm]{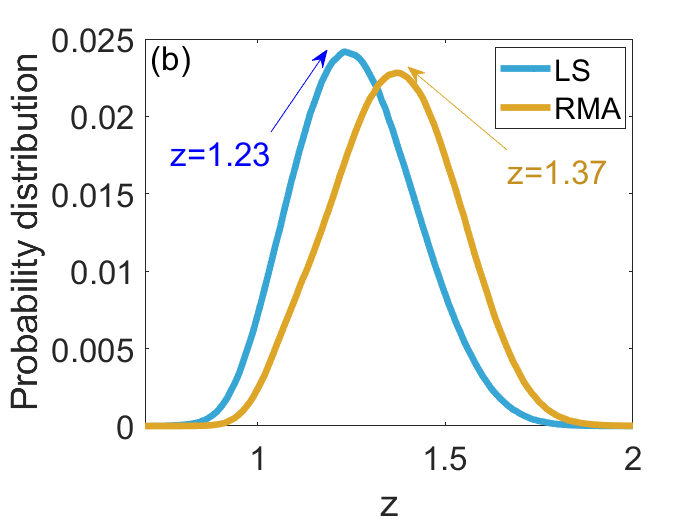}
\\
\end{center}
\caption{Mixtures of simulation data for $N=500$, $0.1\leq\eta\leq 0.5$.  {\bf (a)} Correlation time vs length for data showing LS and RMA fittings to straight lines for a mixture of $\beta_0(N;\eta)$ and $\beta_c(N;\eta)$ data: the corresponding dynamical critical exponents are $z_\text{LS}=1.24\pm 0.11$ and $z_\text{RMA}=1.37\pm 0.10$.   {\bf (b)} Probability distribution of the LS (blue) and RMA (orange) critical exponent $z$. }
 \label{fig13}
\end{figure}
\end{center}\end{widetext}

\section{Discussion}\label{sec:5}
For finite $N$, the HCVM scale-free-chaos phase transition has an extended criticality region in the noise-confinement phase plane bounded by two critical scale-free lines $\beta_0(N;\eta)$ and $\beta_c(N;\eta)$, which separate nonchaotic-chaotic attractors and single-multicluster chaos, respectively. On these lines, macroscopic quantities exhibit power laws in the control parameter (confinement or perception range) and in noise. We compute critical exponents by exploiting these power laws, either by fixing noise and increasing $N$ or by decreasing noise at fixed $N$. 

Power laws for natural midge swarms are obtained using data from different number of insects and species under variable environmental conditions \cite{att14,cav17,cav23}. We mimic these conditions by using a mixture of $\eta$ and $N$ values on the critical lines $\beta_{0,c}(N;\eta)$ of the extended criticality region of the HCVM scale-free-chaos phase transition (Region III of Fig.~\ref{fig1}(c)). Applying to our numerical simulations the same tests used to extract critical exponents from observations, we predict static and dynamic critical exponents in agreement with those observed in natural swarms within the uncertainty range of the data. Furthermore, observed {\em qualitative features} such as the collapse of the NDCCF only at short scaled times $t/\xi^z$ \cite{cav17} or the swarm shape in Figures \ref{fig2}(b) and \ref{fig2}(c) (condensed core surrounded by insect vapor \cite{sin17}) agree with HCVM simulations and with HCVM mean-field theory \cite{gon23mf}. Collapse of $g(t)$ only for short scaled times suggests \cite{gon23} that several time scales are involved in HCVM simulation data near the scale-free-chaos phase transition and in measurements on natural swarms \cite{cav17}.

Contrastingly, numerical simulations near ordering scale-free transitions produce collapse of the NDCCF for all scale times $k_c^zt$; see Figures~2(c) and 2(d) of \cite{cav17} for the periodic VM and \cite{cav23} (Fig.~20 of the Supplementary Material) for the ISM, also with periodic boundary conditions. Collapse of the NDCCF for all scaled times indicates that a single correlation time is involved in the ordering phase transition of these models. Despite being models of active matter far from equilibrium, the ordering phase transition involves spatially homogeneous phases that are invariant under translations. This facilitates using RG theory to calculate critical exponents of the active E/F and G models  as a control parameter tends to the critical point of the transition \cite{cav23}. In fact, in the absence of noise, phases are constant solutions of the governing equations of the model, which are nonlinear stochastic partial differential equations with space independent coefficients. RG calculations are based upon perturbation theory about these simple phases \cite{wil74,hoh77,ami05,cav23}. The theories based on the ordering phase transition predict accurately the dynamical critical exponent ($z=1.35$), but fail to predict the static critical exponents (they predict $\nu=0.748$, $\gamma=1.171$ instead of the observed values $\nu=0.35\pm 0.10$ and $\gamma=0.9\pm 0.2$ \cite{att14}) \cite{cav23}, the limited collapse of the NDCCF \cite{cav17}, or the shape of the swarm \cite{att14,sin17}. To belong to the same universality class, theories and experiments should yield the same critical exponents. A satisfactory explanation of insect swarms should reproduce qualitative features such as the shape of the swarm and the limited collapse of the NDCCF. Inasmuch as the symmetries of the ordering transition do not respond to the qualitative features observed in insect swarms (markers, collapse of NDCCF only for scaled times on a finite interval, etc.), we conclude that the ordering transition between homogeneous phases does not belong to the hypothetical university class of insect swarms. Since it describes qualitative features and provides static and dynamical critical exponents close to measured ones, we think our scale-free-chaos phase transition has a better chance to describe natural swarms. 
 
To check whether natural swarms are close to a scale-free-chaos phase transition, time series from measurements should be used to calculate the largest Lyapunov exponent \cite{gon23}. If the data are not sufficient, calculating the scale-dependent Lyapunov exponent could test whether noisy chaos is consistent with observations \cite{gon23}. Random motion observed in experiments \cite{ni15,rey16} may point out to midge swarms being in the vicinity of chaotic attractors.

From a theoretical standpoint, a future RG theory of the HCVM would ascertain the class of universality of its scale-free-chaos phase transition. Numerical simulations indicate that the RG flow should include a line of critical points comprising the point of zero noise and confinement in Fig.~\ref{fig1} \cite{gon23,gon23mf}. This feature is absent from the RG flow of the ordering transition of active stochastic PDEs \cite{cav23}. 

Coming back to the question on whether biological systems are close to criticality \cite{mor11}, our results may introduce a new twist to the analogy with phase transitions. Sometimes it is difficult to identify a control parameter of the biological systems and ascertain how far they are from the critical point. In particular, this is the case if the measurements involve observations under different external conditions or the data refer to systems comprising different numbers of entities \cite{att14,cav17,cav23}. In these cases, a mixture of data over the critical region of a given theory may explain observations, as we endeavor to show here for natural midge swarms.

\begin{acknowledgments}
This work has been supported by the FEDER/Ministerio de Ciencia, Innovaci\'on y Universidades -- Agencia Estatal de Investigaci\'on grants PID2020-112796RB-C21 (RGA) and PID2020-112796RB-C22 (LLB), by the Madrid Government (Comunidad de Madrid-Spain) under the Multiannual Agreement with UC3M in the line of Excellence of University Professors (EPUC3M23), and in the context of the V PRICIT (Regional Programme of Research and Technological Innovation). RGA acknowledges support from the Ministerio de Econom\'\i a y Competitividad of Spain through the  Formaci\'on de Doctores program Grant PRE2018-083807 cofinanced by the European Social Fund.
\end{acknowledgments}

\appendix
\section{Chaotic and noisy dynamics (adapted from \cite{gon23})}\label{ap:a}
We calculate the LLE in different ways that are complementary to each other: (i) directly from the equations by using the Benettin {\em et al} algorithm (BA) \cite{ben80}, and from time traces of the center-of-mass motion or the NDCCF to reconstruct the phase space of the chaotic attractor by means of: (ii) the scale-dependent Lyapunov exponent (SDLE) algorithm \cite{gao06} and (iii) the Gao-Zheng algorithm \cite{gao94}. Using the BA requires knowing the equations of the model whereas time traces can be obtained from numerical simulations of equations or from experiments and observations. The SDLE algorithm is useful to separate the cases of mostly deterministic chaos from noisy chaos and mostly noise even in the presence of scarce data and a reconstruction of the attractor that is not very precise \cite{gao06} whereas the Gao-Zheng algorithm requires more data points \cite{gao94}. We now describe these different algorithms and illustrate the results they provide for the HCVM. In all cases, we eliminate the effects of initial conditions by leaving out the first 30000 time steps before processing the time traces.

\subsection{Benettin algorithm}
We have to simultaneously  solve Eqs.~\eqref{eq1} and the linearized system of equations
\begin{widetext}
\begin{eqnarray}
\delta\mathbf{\tilde{x}}_i(t+1)\!&=&\! \delta\mathbf{\tilde{x}}_i(t)+\delta\mathbf{\tilde{v}}_i(t+1),  \quad i=1,\ldots, N,  \label{eqb1}\\
\delta\mathbf{\tilde{v}}_i(t+1)\!&=&\! v_0\mathcal{R}_\eta\!\left[\!\left(\mathbb{I}_3-\frac{[\sum_{|\mathbf{x}_j-\mathbf{x}_i|<R_0}\mathbf{v}_j(t)-\beta\mathbf{x}_i(t)]^T[\sum_{|\mathbf{x}_j-\mathbf{x}_i|<R_0}\mathbf{v}_j(t)-\beta\mathbf{x}_i(t)]}{|\sum_{|\mathbf{x}_j-\mathbf{x}_i|<R_0}\mathbf{v}_j(t)-\beta\mathbf{x}_i(t)|^2}\right)\!\cdot \frac{\sum_{|\mathbf{x}_j-\mathbf{x}_i|<R_0}\delta\mathbf{\tilde{v}}_j(t)-\beta\delta\mathbf{\tilde{x}}_i(t)}{|\sum_{|\mathbf{x}_j-\mathbf{x}_i|<R_0}\mathbf{v}_j(t)-\beta\mathbf{x}_i(t)|}\right]\!,  \nonumber
\end{eqnarray}
\end{widetext}
in such a way that the random realizations $\mathcal{R}_\eta$ are exactly the same for Eqs.~\eqref{eq1} and \eqref{eqb1}. The initial conditions for the disturbances, $\delta\mathbf{\tilde{x}}_i(0)$ and $\delta\mathbf{\tilde{v}}_i(0)$, can be randomly selected so that the overall length of the vector $\delta\bm{\chi}=(\delta\mathbf{\tilde{x}}_1,\ldots,\delta\mathbf{\tilde{x}}_N,\delta\mathbf{\tilde{v}}_1,\ldots,\delta\mathbf{\tilde{v}}_N)$ equals 1. After each time step $t$, the vector $\delta\bm{\chi}(t)$ has length $\alpha_t$. At that time, we renormalize $\delta\bm{\chi}(t)$ to $\hat{\bm{\chi}}(t)=\delta\bm{\chi}(t)/\alpha_t$ and use this value as initial condition to calculate $\delta\bm{\chi}(t+1)$. With all the values $\alpha_t$ and for sufficiently large $l$, we calculate the Lyapunov exponent as
\begin{eqnarray}
&&\lambda_1= \frac{1}{l}\sum_{t=1}^l\ln\alpha_t, \label{eqb2}\\
&& \alpha_t=|\delta\bm{\chi}(t)|=|(\delta\mathbf{\tilde{x}}_1(t),\ldots,\delta\mathbf{\tilde{x}}_N(t),\delta\mathbf{\tilde{v}}_1(t),\ldots,\delta\mathbf{\tilde{v}}_N(t))|, \nonumber
\end{eqnarray}
See Figures 17 and 18 of ~\cite{gon23} for convergence of the BA.

\begin{widetext}
\begin{center}
\begin{figure}[ht]
\begin{center}
\includegraphics[width=5.8cm]{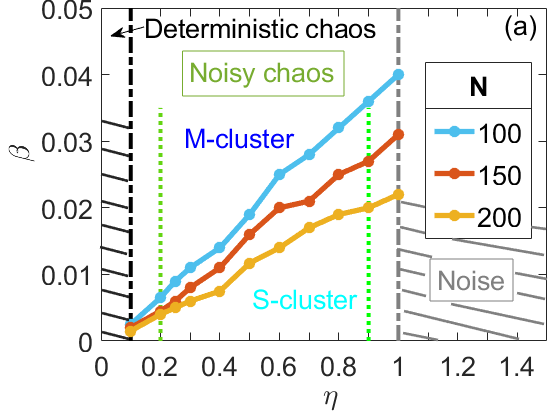}
\includegraphics[width=5.8cm]{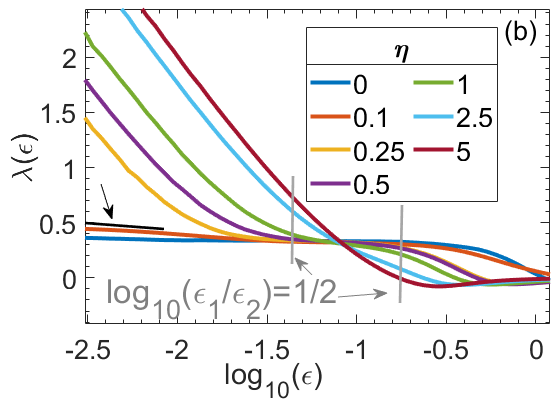}
\includegraphics[width=6.1cm]{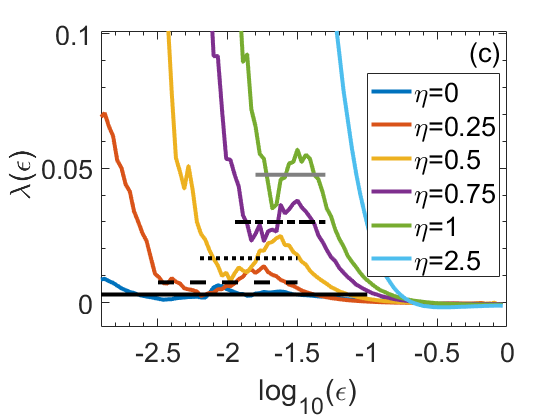}
\end{center}
\caption{{\bf Scale-free chaos. (a)} Phase diagram $\beta$ vs $\eta$ exhibiting regions of deterministic and noisy chaos, and of noisy disorder. The vertical lines at $\eta=0.2$ and 0.9 correspond to the maximum correlation length observed in experiments and to the noise for which the dynamic correlation function ceases to be flat near $t=0$, respectively. Noise swamps chaos for $\eta\geq1$. The three lines of critical points in the noisy chaos region correspond to critical confinement $\beta_c(N,\eta)$ for $N=100, 150, 200$.  They separate multicluster (M-cluster) from single cluster (S-cluster) chaos. {\bf (b)} Largest scale-dependent Lyapunov exponent as a function of the scale parameter $\epsilon$ for $N=100$, different values of $\eta$, two lagged coordinates $m=2$ and $\beta=\beta_c(N,\eta)$. The LLE is the value of $\lambda(\varepsilon)$ at a plateau $(\varepsilon_1,\varepsilon_2)$ whose width satisfies $\log_{10}(\varepsilon_2/\varepsilon_1)\geq 1/2$. The vertical lines mark the width of the critical plateau at which: $\log_{10}(\varepsilon_2/\varepsilon_1)= 1/2$ and correspond to the vertical dot-dashed lines in Panel (a). The black line and arrow mark the very small slope of the SDLE for noise values close to deterministic chaos. By convention \cite{gao06}, noise swamps chaos when $\log_{10}(\varepsilon_2/\varepsilon_1)< 1/2$. {\bf (c)}  Largest scale-dependent Lyapunov exponent as a function of the scale parameter $\varepsilon$ for $N=100$, different values of $\eta$, and $\beta=\beta_c(N,\eta)$ with $m=6$, instead of $m=2$ as in Panel (b). The averages of the oscillations corresponding to the plateau region in Panel (b) increase with the noise $\eta$ indicating that so does the LLE: $\lambda_1(0)\sim 0.003, \lambda_1(0.25)\sim 0.0075, \lambda_1(0.5)\sim 0.0165, \lambda_1(0.75)\sim 0.03, \lambda_1(1)\sim 0.0476$. Reproduced from Figure 3 of \cite{gon23}.} 
 \label{fig14}
\end{figure}
\end{center}
\end{widetext}

\subsection{Scale dependent Lyapunov exponents}
We use scale dependent Lyapunov exponents (SDLE) from the CM motion to characterize deterministic and noisy chaos as different from noise \cite{gao06}. 

Adding the components of $\mathbf{X}(t)$, we form the time series $x(t)=X_1(t)+X_2(t)+X_3(t)$. To calculate the SDLE, we construct the lagged vectors: $\mathbf{X}_\alpha=[x(\alpha),x(\alpha+\tilde{\tau}),..,x(\alpha+(m-1)\tilde{\tau})]$. The simplest choice is $m=2$ and $\tilde{\tau}=1$ (other values can be used, see below). From this dataset, we determine the maximum $\varepsilon_\text{max}$ and the minimum $\varepsilon_\text{min}$ of the distances between two vectors, $\|\mathbf{X}_\alpha-\mathbf{X}_\beta\|$. Our data is confined in $[\varepsilon_\text{min},\varepsilon_\text{max}]$. Let $\varepsilon_0$, $\varepsilon_t$ and $\varepsilon_{t+\Delta t}$ be the average separation between nearby trajectories at times 0, $t$, and $t+\Delta t$, respectively. The SDLE is
 \begin{subequations}\label{eqb3}
 \begin{eqnarray}
\lambda(\varepsilon_t)=\frac{\ln\varepsilon_{t+\Delta t}-\ln\varepsilon_t}{\Delta t}. \label{eqb3a}
\end{eqnarray}
The smallest possible $\Delta t$ is of course the time step $\tilde{\tau}=1$, but $\Delta t$ may also be chosen as an integer larger than 1. Gao {\em et al} introduced the following scheme to compute the SDLE \cite{gao06}. Find all the pairs of vectors in the phase space whose distances are initially within a shell of radius $\epsilon_k$ and width $\Delta\epsilon_k$:
\begin{eqnarray}
\varepsilon_k\leq\|\mathbf{X}_\alpha-\mathbf{X}_\beta\|\leq\varepsilon_k+\Delta\varepsilon_k,\quad k=1,2,\ldots.\label{eqb3b}
\end{eqnarray}
We calculate the Lyapunov exponent \eqref{eqb3a} as follows:  
\begin{eqnarray*}
&&\lambda(\varepsilon_t)=
\frac{\langle\ln\|X_{\alpha+t+\triangle t}-X_{\beta+t+\triangle t}\|-\ln\|X_{\alpha+t}-X_{\beta+t}\|\rangle_{k}}{\triangle t},\quad \quad 
\end{eqnarray*} 
where $\langle\rangle_{k}$ is the average within the shell $(\varepsilon_k,\varepsilon_k+\triangle \varepsilon_k)$. The shell dependent SDLE $\lambda(\varepsilon)$ in Fig.~\ref{fig14}(b) displays the dynamics at different scales for $\tilde{\tau}=1$ and $m=2$ \cite{gao06}. Using two lagged coordinates produces plateaus having a value of $\lambda(\varepsilon)$ equal to the LLE of deterministic chaos. This value differs from the LLE calculated using the BA or a more appropriate reconstruction of the phase space involving more lagged coordinates (see below). However, the SDLE with $m=2$ yields a qualitative idea of the effects of noise on chaos. In deterministic chaos, $\lambda(\varepsilon)>0$ presents a plateau with ends $\varepsilon_1< \varepsilon_2\ll 1$, in noisy chaos, this plateau is preceded and succeeded by regions in which $\lambda(\varepsilon)$ decays as $-\gamma\ln\varepsilon$, whereas it shrinks and disappears when noise swamps chaos. As $\eta$ increases, $\lambda(\varepsilon)$ first decays to a plateau for $\eta=0.1$. A criterion to distinguish (deterministic or noisy) chaos from noise is to accept the largest Lyapunov exponent as the positive value at a plateau $(\varepsilon_1,\varepsilon_2)$ satisfying 
\begin{eqnarray}
\log_{10}\frac{\varepsilon_2}{\varepsilon_1}\geq \frac{1}{2}.  \label{eqb3d}
\end{eqnarray}\end{subequations}
For $\eta=0.5$, the region where $\log_{10}(\varepsilon_2/\varepsilon_1)= 1/2$ is marked in Fig.~\ref{fig14}(b) by vertical lines. Plateaus with smaller values of $\log_{10}(\varepsilon_2/ \varepsilon_1)$ or their absence indicate noisy dynamics \cite{gao06}. This occurs for $\eta=1$. The ends of the interval $(0.1,1)$ of noisy chaos are marked by two vertical dashed lines in Fig.~\ref{fig14}(a). 

\subsection{Largest Lyapunov exponent from high dimensional reconstructions of CM motion}
The previous reconstruction of the phase space for CM motion used to calculate SDLE considers 2D lagged vectors ($m=2$). This produces useful qualitative phase diagrams with flat plateaus, but the dimension of this vector space is too small to reconstruct faithfully the attractor. More realistic CM trajectories in higher dimension contain self-intersections in dimension 2. This explains the different values of the LLE found in the SDLE plateaus of Fig.~\ref{fig14}(b) as compared with those found by the BA of Eq.~\eqref{eqb2}. To reconstruct  safely a chaotic attractor, the dimension of the lagged vectors should surpass twice the fractal dimension $D_0$ \cite{ott93}. For the HCVM, $m=6$ is sufficient in view of Figure 6 of Ref.~\cite{gon23}. However, the SDLE $\lambda(\varepsilon)$ presents oscillations as indicated in Fig.~\ref{fig14}(c) and their average values replace the plateaus in Fig.~\ref{fig14}(b). In contrast with  Fig.~\ref{fig14}(b), the averaged oscillations produce LLEs that increase with noise. Averaging oscillations is not going to produce precise values of the LLE. Thus, we calculate the LLE from the lagged coordinates with $m=6$ using the Gao-Zheng algorithm \cite{gao94}. This requires constructing the quantity $\Lambda(k)$ whose slope near the origin gives the LLE \cite{gao94}
\begin{eqnarray}
\Lambda(k)= \left\langle\ln\frac{\|X_{i+k}-X_{j+k}\|}{\|X_i-X_j\|}\right\rangle.\label{eqb4}
\end{eqnarray}
Here the brackets indicate ensemble average over all vector pairs with $\|X_i-X_j\|<r^*$ for an appropriately selected small distance $r^*$. Figure 20 of \cite{gon23} displays the graph of $\Lambda(k)$ given by Eq.~\ref{eqb4}. The slopes of $\Lambda(k)$ for different values of $N$ at $\beta_c(N)$ equal the LLEs, increase with $\beta$ and agree with the averaged oscillations marked in Fig.~\ref{fig14}(c).

\end{document}